\documentclass[fleqn,usenatbib]{mnras}
\usepackage{amsmath}
\pdfoutput=1
\usepackage{txfonts}
\usepackage[table,xcdraw]{xcolor}
\usepackage{longtable}

\usepackage[T1]{fontenc}
\usepackage{ae,aecompl}

\usepackage{graphicx}	
\usepackage{caption}

\defcitealias{carlsten20a}{Carlsten et. al. (2020a)}
\defcitealias{carlsten20b}{Carlsten et. al. (2020b)}
\defcitealias{zanatta21}{Z21}
\defcitealias{denbrok14}{dB14}
\defcitealias{sanchez-janssen19b}{SJ19}

\title[NSCs in the Shapley Supercluster]{NSCs from groups to clusters: A catalogue of dwarf galaxies in the Shapley Supercluster and the role of environment in galaxy nucleation}

\author[Zanatta et al.]{
Em\'ilio Zanatta$^{1}$\thanks{E-mail: emiliojbzanatta@gmail.com},
Rub\'en S\'anchez-Janssen$^{2}$\thanks{E-mail: ruben.sanchez-janssen@stfc.ac.uk},
Rafael S. de Souza$^{3}$, \newauthor
Ana L. Chies-Santos$^{4}$,
John P. Blakeslee$^{5}$
\\
$^{1}$Universidade de São Paulo, IAG, Rua do Matão 1226, Cidade Universitária, São Paulo 05508-900, Brazil\\
$^{2}$STFC UK Astronomy Technology Centre, Royal Observatory, Blackford Hill, Edinburgh, EH9 3HJ, UK\\
$^{3}$Centre for Astrophysics Research, University of Hertfordshire, College Lane, Hatfield, AL10~9AB, UK\\
$^{4}$Instituto de F\'isica, Universidade Federal do Rio Grande do Sul, Porto Alegre, R.S., Brazil\\
$^{5}$NSF’s NOIRLab, 950 N. Cherry Ave., Tucson, AZ 85719, USA
}

\date{Accepted XXX. Received YYY; in original form ZZZ}

\pubyear{2022}

\begin{document}
\label{firstpage}
\pagerange{\pageref{firstpage}--\pageref{lastpage}}
\maketitle


\begin{abstract}
Nuclear star clusters (NSCs) are dense star clusters located at the centre of galaxies spanning a wide range of masses and morphologies. Analysing NSC occupation statistics in different environments provides an invaluable window into investigating early conditions of high-density star formation and mass assembly in clusters and group galaxies. We use \textit{HST/ACS} deep imaging to obtain a catalogue of dwarf galaxies in two galaxy clusters in the Shapley Supercluster: the central cluster Abell 3558 and the northern Abell 1736a. The Shapley region is an ideal laboratory to study nucleation as it stands as the highest mass concentration in the nearby Universe. We investigate the NSC occurrence in quiescent dwarf galaxies as faint as $M_{I} = -10$ mag and compare it with all other environments where nucleation data is available. We use galaxy cluster/group halo mass as a proxy for the environment and employ a Bayesian logistic regression framework to model the nucleation fraction ($f_{n}$) as a function of galaxy luminosity and environment. We find a notably high $f_n$ in Abell 3558: at $M_{I} \approx -13.1$ mag, half the galaxies in the cluster host NSCs. This is higher than in the Virgo and Fornax clusters but comparable to the Coma Cluster. On the other hand, the $f_n$ in Abell 1736a is relatively lower, comparable to groups in the Local Volume. We find that the probability of nucleation varies with galaxy luminosity remarkably similarly in galaxy clusters. These results reinforce previous findings of the important role of the environment in NSC formation/growth. 
\end{abstract}




\section{Introduction}

Nuclear star clusters (NSCs) are generally identified as luminous overdensities at the very central regions of galaxies. They are a class of extremely dense and bright objects, with masses and half-light radii ranging from $10^4$ to $10^8 \mathcal{M}_{\odot}$ and from 1 to 50 pc, respectively \citep{sanchez-janssen19b, neumayer20}. They are commonly found in galaxies of a wide range of masses, morphologies and environments, but are not ubiquitously present \citep{boker02a, seth06, georgiev09, georgiev14, neumayer20}. It has been established that the fraction of galaxies that host NSCs increases universally as a function of galaxy luminosity or stellar mass, with a peak at $\sim 10^9 \mathcal{M}\odot$ \citep{denbrok14, munoz15, sanchez-janssen19b, neumayer20, hoyer21, su22}, then declining for both higher \citep{cote06, turner12, baldassare14a} and lower \citep{denbrok14, ordenes-briceno18, sanchez-janssen19b} galaxy masses. Despite their common presence, the question of what processes are involved in the formation and growth of NSCs is still under debate. Nevertheless, to study how exactly these objects form and evolve is essential for the discussion of how the earlier events of star formation, and the subsequent formation of the earliest star clusters, unfold in galaxies \citep{baldassare14a, sanchez-janssen19b, neumayer20, leaman21, fahrion22}.

NSC formation is generally thought to derive from two, non-exclusive mechanisms. One is the dissipationless decay of globular clusters (GCs) to the centre of the host galaxy gravitational potential, where eventually star clusters massive enough to survive effects of dynamical friction merge into a larger and denser object \citep{tremaine75,arca-sedda14,gnedin14}. Another process comprises the \textit{in-situ} increase in star formation at the centre of galaxies credited to the inflow of gas \citep{bekki03, bekki04, antonini13}. The first is credited to be the main driver of NSC formation at the low mass regime ($\lesssim 10^9 \mathcal{M}\odot$) \citep{neumayer20, fahrion22} and explains the metal-poor stellar populations observed in nearby NSCs \citep{alfaro-cuello19a, johnston20, fahrion20}, while the latter can explain the presence of young and metal-rich stellar populations observed in NSCs of late-type galaxies \citep{carson15, kacharov18, nguyen18}. A combination of both process is likely required to explain the full range of observed NSC properties, such as stellar populations, morphology and kinematics \citep{hartmann11, antonini13, antonini15, guillard16}.    

Remarkably, recent studies have observed that the nucleation fraction of dwarf galaxies ($\mathcal{M}_\odot \lesssim  10^9$) display a strong dependence with the environment, which does not arise naturally from the aforementioned NSC formation scenarios \citep{sanchez-janssen19b, zanatta21, hoyer21, carlsten22a, su22}. \citet{sanchez-janssen19b}, hereafter \citetalias{sanchez-janssen19b}, observed a high nucleation fraction in dwarfs at the Coma cluster, followed by the Virgo and Fornax clusters, with the lowest nucleation fraction found in satellites in groups at the Local Volume. One important caveat of this analysis was the limited magnitude range of the Coma cluster sample from \citet{denbrok14} (hereafter, \citetalias{denbrok14}), which comprised a significantly brighter sample ($M_I \geq -13$) than the other environments analysed ($M_I \geq -9$). Using deep Hubble Space Telescope/Advanced Camera for Surveys (HST/ACS) imaging of the Coma Cluster, \citet{zanatta21} (hereafter \citetalias{zanatta21}) addressed this issue presenting nucleation information for dwarfs at the faint end of Coma ($M_I \geq -9$) and confirming that NSC occurrence is systematically higher for galaxies located in more massive environments. In such work, it has been shown that a $M_I = -15$ mag dwarf has an $\sim$ 80\% probability of being nucleated in Coma, compared to the $\sim$ 50\% for quiescent satellites in groups at Local Volume. The same conclusions have been obtained subsequently in works such as \citet{hoyer21} and \citet{carlsten22a}, that confirmed the lower nucleation in Local Volume dwarfs compared to galaxy clusters, as well as in \citet{su22}, that shows the higher nucleation fraction in the Fornax main cluster when compared to its neighbouring Fornax A group. 

In this work we seek to further investigate the role of mass and environment as major drivers of NSC formation. We use the same procedure from \citetalias{zanatta21} to a set of HST images of prominent clusters within the Shapley Supercluster (SSC), namely Abell 3558 (hereafter, A3558), at its very centre, and Abell 1736a (hereafter A1736a), to the north. SSC is by far the largest concentration of mass in the local Universe \citep{haines18a, higuchi20}, located at the distance of $\sim$ 150-200 Mpc and widely regarded as the most likely candidate to be responsible for the phenomenon known as the \textit{Great Attractor}, i.e., the mass concentration to where all galaxies in the local group are moving towards. The SSC core is a roughly spherical region with the A3558 cluster at its very centre \citep{quintana20}. A3558 has an estimated peak density of $\approx$ 160 galaxies per Mpc$^2$ \citep{haines18a, higuchi20}, presenting an ideal case to investigate the nucleation phenomenon in an extremely dense environment. A1736a provides a second window into an environment within the SSC, and in a slightly different density regime, as this cluster is relatively poor, with an estimated peak density of $\approx$ 31 galaxies per Mpc$^2$ \citep{higuchi20}. Furthermore, we will compare the nucleation in A3558 and A1736a with literature data from galaxy clusters (Coma, Virgo and Fornax) to Local Volume groups, effectively covering two decades of environment halo mass. 

This paper is structured as follows. In Section \ref{sec:data} we present the data used in this work, followed by a detailed description of the galaxy and NSC detection and measurement techniques in Section \ref{sec:phot}. In Section \ref{sec:logit} we present the statistical methodology developed to infer the nucleation fraction in A3558 and A1736a, which are compared to that of other environments in Section \ref{sec:results}. In Section \ref{sec:discussion} we discuss the main results within the context of observational and theoretical work on the galaxy nucleation fraction. Finally, in Section \ref{sec:conclusion} we summarise our results and their significance for NSC formation scenarios. 

Throughout this work we adopt $H_0 = $70 kms$^{-1}$ Mpc $^{-1}$, $\Omega_m =$ 0.3 and $\Omega_\Lambda =$ 0.7. We adopt a redshift to A3558 of $z = 0.0477$, which corresponds to a physical scale of 0.936 kpc/arcsec, a distance modulus of $(m-M) = 36.52$ mag and a distance of 211.58 Mpc. For A1736a we adopt a redshift of $z = 0.0350$, which corresponds to a physical scale of 0.697 kpc/arcsec, a distance modulus of $(m-M) = 35.89$ mag and a distance of 148.8 Mpc \citep{blakeslee08}.  

\begin{figure*}
    \centering
    \includegraphics[width=\textwidth]{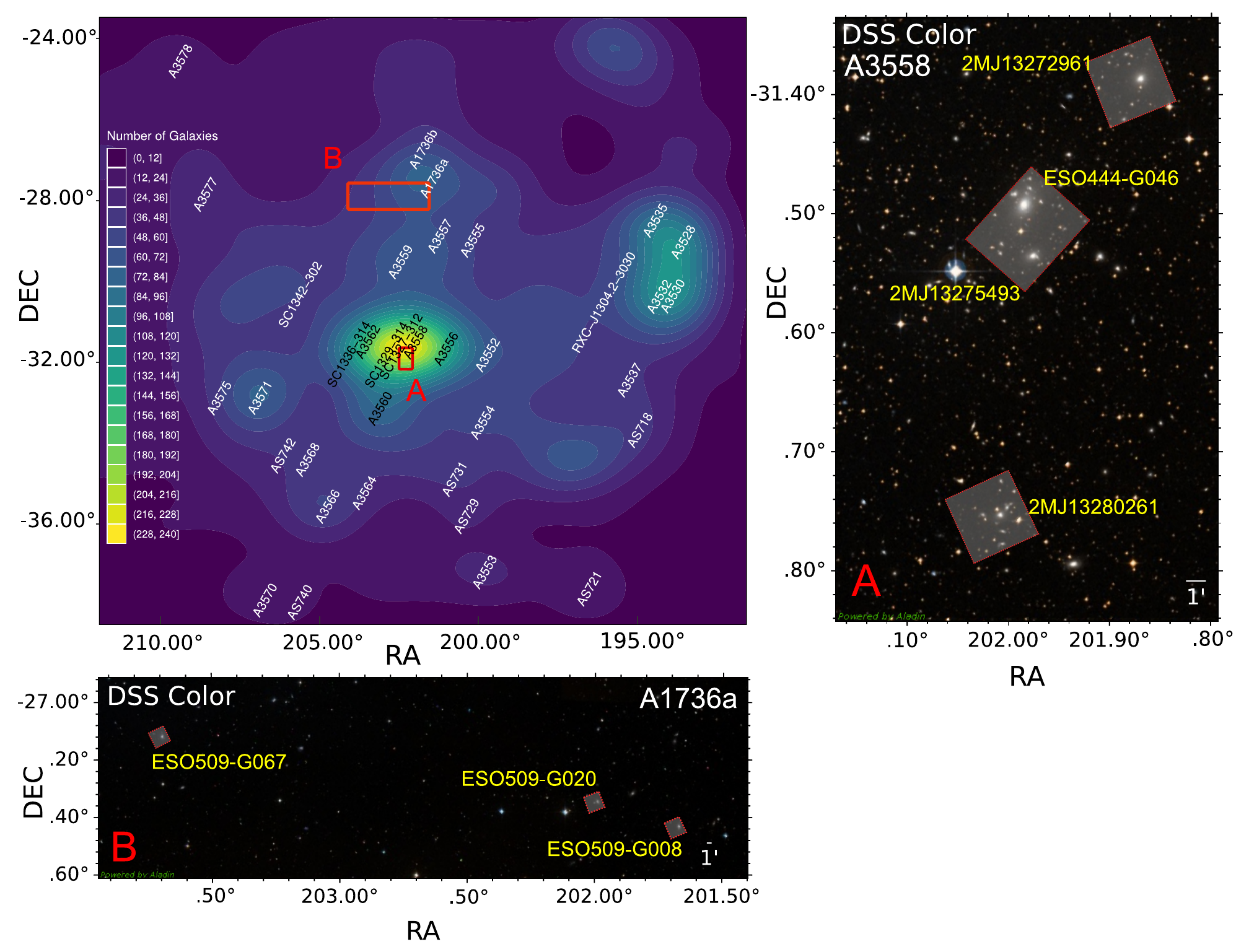}
    \caption{Location of the clusters studied in this work in the context of the Shapley supercluster region in the sky. \textit{Top right, larger panel:} Shapley region with colours denoting the number of galaxies, using the data from \citet{quintana20}. Labelled are a selection of important galaxy clusters associated with Shapley, with red rectangles denoting the approximated area covered by the right (A) and bottom (B) panels, around the location of the clusters studied in this work, i.e., A3558 and A1736a. \textit{Right panel:} DSS2 colour-composite zoom into the region of the A3558 cluster where the four HST fields used in this work are located, represented in this figure by the white regions with red contours. The yellow labels indicate the brightest galaxy in each field, also shown in Table \ref{tab:data}. Notice that two out of the four fields in this cluster are contiguous. \textit{Bottom panel:} Same as panel A, but for the area around A1736a, at the northern region of Shapley. Notice the different scales of the panels, indicated by the scale bar at the lower right of each panel.}
    \label{fig:fields}
\end{figure*}

\begin{figure*}
    \centering
    \includegraphics[width=\textwidth]{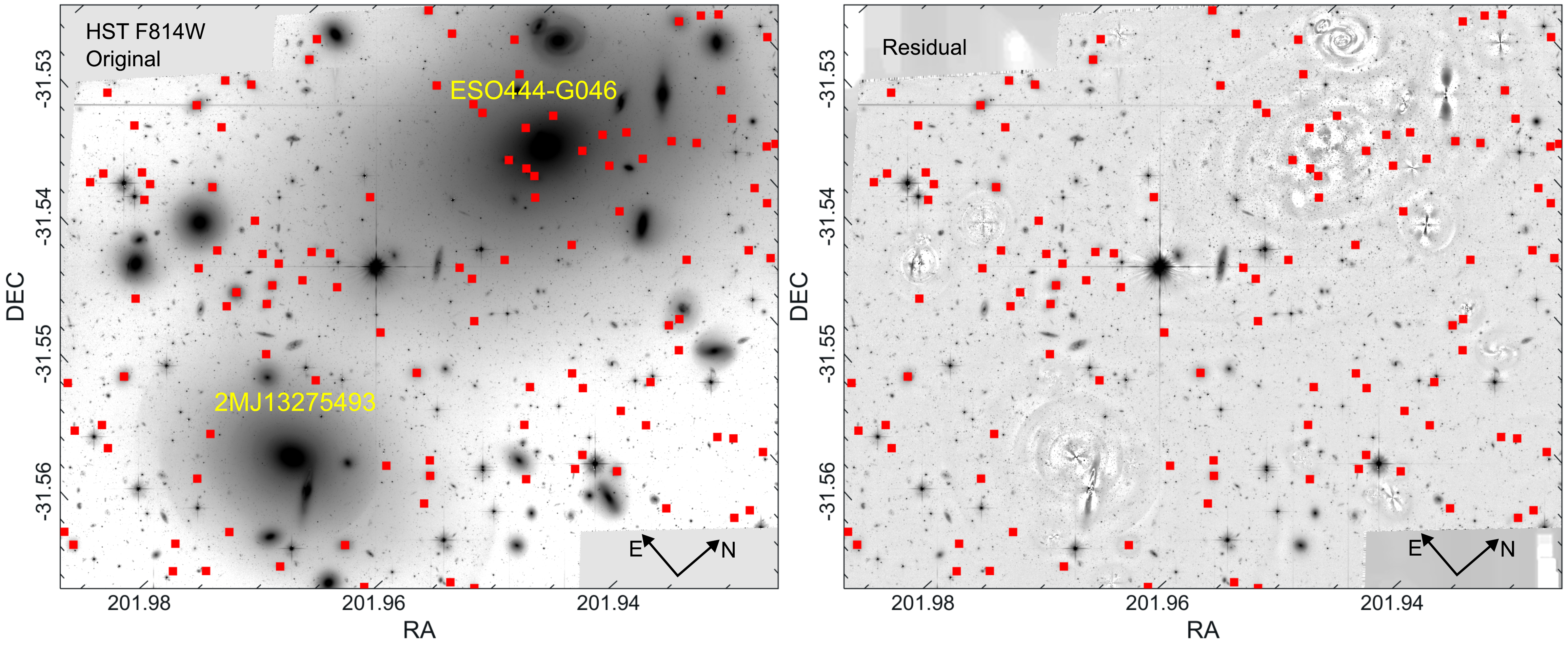}
    \caption{Example of an original and a residual bright galaxy subtracted image at the left and right panels, respectively. This treatment was necessary to improve the detections of faint galaxies in all fields analysed. In this example, we show in both panels only the two central and contiguous fields in the A3558 cluster, centered around the galaxies labelled in yellow. We also show in red markers the position of the quiescent dwarf galaxies detected, modelled and presented in Table \ref{tab:catalog}.}
    \label{fig:subgals}
\end{figure*}

\begin{table}
\centering
\caption{Data observed as part of the HST/ACS program GO-10429 to observe the core of the Shapley Supercluster. From left to right: central galaxy in the field, associated cluster in the Abell catalogue, absolute V-band magnitude from NED, redshift also from NED and exposure time in seconds from the HST/ACS observations. The shaded rows indicate the observations that comprise the images used in this work and shown in Figure \ref{fig:fields}.}
\label{tab:data}
\resizebox{\columnwidth}{!}{%
\begin{tabular}{lllll}
\hline
Galaxy & Cluster & $M_V$ & z & Exp. Time \\ 
& & & & (seconds) \\ \hline
\rowcolor[HTML]{DDDDDD}
ESO509-G008 & A1736a &-22.8 &0.0350 &18567 \\
\rowcolor[HTML]{DDDDDD}
ESO509-G020 & A1736a &-22.5 &0.0350 &18567 \\
\rowcolor[HTML]{DDDDDD}
ESO509-G067 &A1736a &-22.7 &0.0350 &18567 \\
ESO325-G016 &A3570  &-22.0 &0.0379 &18882 \\
ESO325-G004 &AS0740  &-22.8 &0.0350 &18882 \\
ESO383-G076 &A3571  &-23.6 &0.0397 &21081 \\
2masxj13481399-3322547 & A3571  & -21.9 & 0.0397 & 21081 \\
\rowcolor[HTML]{DDDDDD}
ESO444-G046 &A3558  &-23.8 &0.0477 &34210\\
\rowcolor[HTML]{DDDDDD}
2masxj13275493-3132187 &A3558  &-22.4 &0.0477 &35550 \\
\rowcolor[HTML]{DDDDDD}
2masxj13272961-3123237 &A3558  &-22.9 &0.0477 &35550 \\
\rowcolor[HTML]{DDDDDD}
2masxj13280261-3145207&A3558  &-22.1 &0.0477 &35550 \\ \hline
\end{tabular}%
}
\end{table}

\section{Data}
\label{sec:data}

The SSC data used in this work were obtained as part of the program GO-10429 (PI: J. Blakeslee) using the Advanced Camera for Surveys Wide Field Channel (ACS/WFC) onboard the Hubble Space Telescope (HST) in January 2005. The program is described in \citet{blakeslee07}. Various galaxies within the SSC were imaged in the F814W ($\approx$~$I$) filter during 114 orbits. The target galaxies, associated clusters and exposure times are shown in Table \ref{tab:data}. Highlighted are the images used for the analysis presented in this work. We use only the images concerning the A3558 and A1736a clusters, for which the program retrieved most data. The remaining images are centered on relatively poor Shapley clusters and our analysis of these resulted in galaxy catalogues with not enough sample size to properly apply the statistical methodology discussed in Section \ref{sec:logit}. In Figure \ref{fig:fields} we show the location of A3558 and A1736a within the context of the SSC. At the top right panel, we show galaxy density contours in the region based on a 2D kernel density estimation and galaxy positions from the \citet{quintana20} catalogue. North is up and East is left. In this panel we can see that A3558 is located at the very peak of galaxy density in the SSC, while A1736a is located at a mild overdensity at the north. In the same figure we show the location of the fields used in this work over DSS2 Color images.  

The data used in this work was previously described in \citet{blakeslee08}, and we refer the reader to such work for details concerning data processing. Briefly, the images were dithered to fill the gap between the two ACS/WFC detectors, followed the standard pipeline processing from the STScI/Mikulski Archive for Space Telescopes (MAST) and had the charge-transfer efficiency (CTE) correction algorithm of \citet{anderson10} applied. Finally, the CTE-corrected exposures were then processed with \textsc{apsis} \citep{blakeslee03} to produce the final corrected images. 

\subsection{NSCs in other environments from the literature}
\label{sec:other}

In addition to the A3558 and A1736a clusters, in this work we also analyse data for dwarf early-type galaxies in the Virgo cluster \citepalias[from][]{sanchez-janssen19b}, the Fornax cluster \citep[from][]{su22}, the Coma cluster \citepalias{denbrok14, zanatta21} and in the Local Volume ($D < 12$ Mpc). The latter comprise two independent galaxy samples: The C22 sample, which is mainly based in the Exploration of Local VolumE Satellites (ELVES) Survey data from \citet{carlsten20a} and \cite{carlsten22a, carlsten22b}; and the H21 sample, based on the catalogue from \citet{hoyer21}, which provides nucleation classification for a sample of galaxies in the Updated Nearby Galaxy Catalogue (UNGC, \citealt{karachentsev13}). While the ELVES catalogue is deeper, the UNGC is the most complete reference for Local Volume galaxies as it is kept continuously updated with new dwarf galaxy detections from the literature.

Only early-type galaxies and dwarfs are considered in the analysis. This is to avoid complications related to both the morphology-density relation \citep{dressler80} as well as the notoriously difficult task of identifying NSCs in star-forming galaxies due to the presence of star formation and obscuration by dust (\citealt{ferrarese20}; \citetalias{sanchez-janssen19b}; \citealt{neumayer20}). We also limit the magnitude range of all catalogues used in this work to be within $-7.5 > M_I > -19.0$, to ensure the observational data is able to reliably be modelled with our statistical methodology as well as have a more precise nucleation classification. This is better detailed in sec. \ref{sec:phot}. Additional information on the literature sources for the data used in this work as well as how we match and select the Local Volume samples is presented in Appendix \ref{appendixA}.  

Furthermore, our main observable used to probe environment differences is the cluster/group estimated halo mass. Other tracers could be employed for this task, such as estimates of local galaxy density or the X-ray luminosity of the intracluster medium. The halo cluster mass stands as our preferred choice as a proxy for environment based on its straightforward interpretation and wide literature availability of estimates for nearby galaxy clusters/groups. However, it is important to consider that this choice hinders our capacity of probing the nucleation environmental dependence in smaller scales, such as a clustercentric radial dependence \citep[see][and references therein for studies of this feature]{ordenes-briceno18, hoyer21, su22}. Our choice of environments permits probing NSC occupation in host haloes with masses ranging from $5\times10^{14}$ in A3558 to $10^{12} \mathcal{M}_{\odot}$ at the Local Volume, enabling us to probe the large scale environmental dependence of galaxy nucleation. Our adopted literature halo mass estimates were obtained from different methods, such as based on Dynamics, Stellar Population Models, X-ray observations and GC kinematics. We refer the reader to each of the individual references indicated in the following paragraph for additional details. In the case of Local Volume environments, in the rest of this work the cluster mass estimate is based on the estimate for the halo mass of individual galaxy systems.

We adopted mass estimates for A3558 and A1736a from \citet{lopes18}. We only could find a single literature reference of halo mass for the Abell 1736 region, with the important caveat that \citet{lopes18} calculate the halo mass for the entire A1736 cluster, i.e., not the clusters A1736a and A1736b in separate. This is to be expected due to the difficulties in obtaining precise mass estimates for neighbouring galaxy clusters in such an overcrowded environment as the SSC. Nevertheless, as our images only account for fields in what is considered to be A1736a, therefore our choice of halo mass for this field is clearly and overestimation. We will take this into consideration when analysing the effects of environment and nucleation in terms of estimated halo masses. When necessary, additional adopted mass estimates come from \citet{lokas03}, \citet{mclaughlin99} and \citet{drinkwater01} for Coma, Virgo and Fornax, respectively. For the Local Volume C22 sample, we derive a mean halo mass using the $log(V_{circ})-log(\mathcal{M}_{200})$ relation from the Illustris TNG100 simulations \citep{pillepich18} and the $V_{circ}$ values in Table 1 of \citetalias{carlsten20a}. Exceptions are: NGC 3115, for which the halo mass estimate comes from \citet{alabi17}; M81, from \citet{karachentsev02}; Cen A, from \citet{vandenbergh00}; M31, which comes from \citet{tamm12} and the Milky Way, from \citet{taylor16}. For the Local Volume H21 sample, we do not perform any analysis involving halo mass estimates due to the information on host galaxies for their dwarf galaxy sample not being accurate enough for our purposes, as mentioned previously.   

\begin{figure*}
    \centering
    \includegraphics[width=\linewidth]{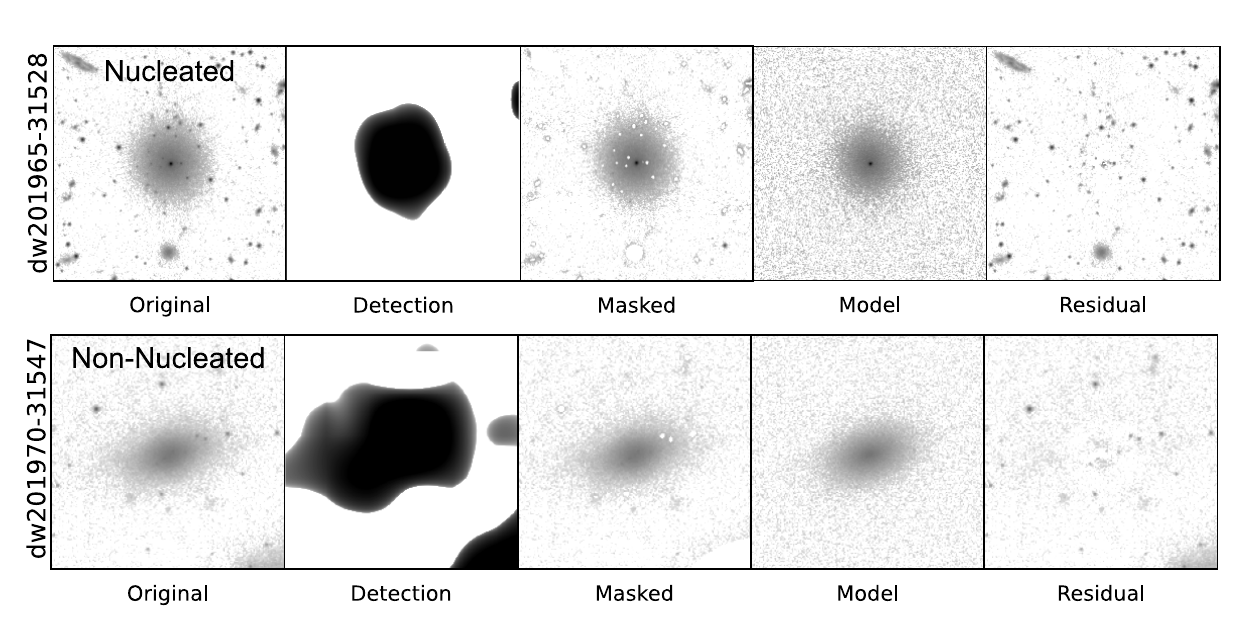}
    \caption{Summary of the procedure to detect and extract photometry for faint galaxies and their NSCs. From left to right: Section of the original image showcasing an example galaxy (nucleated on top, non-nucleated at the bottom). 
    In the second panel, we show the \textsc{SourceExtractor} background image used for galaxy detection. Notice the significant increase in SNR,  which improves the detection limits. 
    The third panel contains the same image as the first panel, but now the point sources detected by the first \textsc{SourceExtractor} are masked--except for the central 6x6 pixels, which are unmasked to reveal the NSC.
    The fourth panel corresponds to the \textsc{galfit} model, to which we have added the typical noise of the $ACS$ images for representation purposes. 
    Finally, in the last panel, we show the residual image from \textsc{galfit} modelling. Both galaxies are presented with the same scaling.
    }
    \label{fig:detectionpanel}
\end{figure*}

\section{Detection, membership and photometry}
\label{sec:phot}

In this work, we use the same technique employed by \citetalias{zanatta21} to detect and analyse faint galaxies in HST images. These methods were build on the ones applied successfully in recent surveys such as \citet{ferrarese20} in Virgo and \citet{eigenthaler18} in Fornax. We refer the reader to \citetalias{zanatta21} for a detailed description, but in the following we briefly describe the method. 

Galaxy detection is executed interactively with an algorithm dedicated to identify low surface brightness objects. Then more than one individual performs a visual inspection of candidates which are selected based on the expected morphological features of quiescent, early-type dwarfs. These comprise basically ellipsoidal shapes and the absence of star formation features, i.e., a smooth surface brightness profile. Objects in the final catalogue are then processed via two-dimensional modelling of the images, resulting in the estimated photometric and structural parameters for the host galaxies and NSCs. 

Our analysis aims to detect NSCs in quiescent dwarfs as in such galaxies the contrast between NSCs and hosts is strong enough to enable clear visual identification. Larger and brighter galaxies often present central components, such as bulges, that may lead to false NSC detections. In the next sections we further discuss the advantages and limitations of this choice, however this does not affect the final goal of this study, since, as we will see later, the final nucleation catalogue presented in this work and those from other environments, gathered from the literature, are all in an homogeneous magnitude range. In a addition to a summary, in the following we detail a few steps in the methodology where it differs from what was done in \citetalias{zanatta21}. 

\subsection{Bright galaxy subtraction}

The extremely crowded environment present in the SSC images used in this work creates a big difficulty to our goal of detecting faint galaxies (see Figure \ref{fig:subgals} for an example in A3558 central fields, but a similar landscape is found for all other fields). Therefore, the first step in our analysis consisted of the modelling and subtraction of the brightest and largest objects to enable an easier detection of faint galaxies.  

We use the task \textsc{ELLIPSE} \citep{ellipse} within the STSDAS package \citep{hanisch92} of the IRAF software \citep{Tody1986} to model and subtract the brightest galaxies in the HST images. The software fits a series of elliptical isophotes of varying centres, ellipticities, orientations, and low-order Fourier terms. The algorithm then interpolates smoothly between the isophotes and extrapolates outward beyond the last. We adopt an iterative approach: We first subtract the brighter galaxies, such as the known brightest cluster galaxies (BCGs) at each field. Then we model large galaxies around the BCGs where visually identify late-type structures, such as bars, disks and spirals. Following this, we model fainter neighbours, remodelling the brighter galaxies with the neighbours subtracted. This process continues until we achieve a clean subtraction of all galaxies that were large enough to have a significant effect on the detection of the low surface brightness objects we aim to study. When we found it significantly hard to achieve flat residuals from the modelling of the BCGs, we complemented the subtraction with a careful background removal with \textsc{SourceExtractor} \citep{bertin96}. We used detection and background parameters individually chosen for each field in order to subtract most of the residuals still present in the residual image but not any haloes of faint galaxies of interest. However, we find that the optimal background parameters in the latter process can still affect haloes of a few large dwarfs with projected positions close to the centre of BCGs. To solve this problem, at the end of the galaxy subtraction step, we output two detection images: one with the BCGs of each field subtracted and nothing else done, and a second one with the subtraction of all large ellipticals, several other large galaxies and the \textsc{SourceExtractor} background subtraction to improve residuals. Each object is modelled in the two images and we select for the final catalogue the \textsc{GALFIT} results with better residuals. In the right panel of Figure \ref{fig:subgals}, we show an example of the final background subtracted image of the central field of A3558.  

\subsection{\textsc{SourceExtractor} detection}

Galaxy detection is carried out in a two-step approach with \textsc{SourceExtractor}, following the methodology presented in \citetalias{zanatta21}. The first run is intended to detect and subtract point sources and generate a detection background image smoothed in a 3x3 grid median filter with mesh sizes of 32 pixels. A second run is then executed on the detection background in which low surface brightness objects are considerably easier to be detected, as it can been attested in the second column of Figure \ref{fig:detectionpanel}, where we show one such detection image. We refer the reader to \citetalias{zanatta21} for details on the \textsc{SourceExtractor} parameters relevant to be fine-tuned in each run. The only difference from the parameters used in that work from this one is that in the first run, this time, we set a minimum detection area of 10 pixels in order to properly detect compact objects such as GCs, foreground stars and background galaxies at the SSC distance. This procedure results in a catalogue of extended object detections in the background image which are then position matched with the original image. Visual inspection to assign cluster membership and nucleation classification is then performed by two of the authors (EZ and RSJ). We classify as members object with smooth and spheroidal morphologies \citep{sanchez-janssen16}, and discard irregular galaxies or those displaying features consistent with ongoing star formation (clumps, arms, bars). We identify NSCs as compact spherical sources that, in projection, lie close to the geometric centre of the candidate galaxy. This is further refined during the galaxy modelling (Section \ref{sec:modelling}). At this point, we find 211 galaxies in A3558 and 58 in A1736a within $14.5 < m_{I} < 26.9$ mag. 

\begin{table*}
\centering
\caption{Photometric and structural parameters for the galaxies detected in A3558 and A1736a obtained using \textsc{GALFIT} using \textsc{SourceExtractor} magnitude and positions as input parameters, as described in the text. From left to right: Identification for each galaxy, right ascension in degrees, declination in degrees, deredenned galaxy magnitude in the $I$ filter, deredenned NSC magnitude in the $I$ filter (when nucleated), S\'ersic index, effective radius in arcseconds, axis ratio and position angle in degrees. Only the 15 brightest objects are shown for each cluster. The full table is available online.}
\label{tab:catalog}
\begin{tabular*}{\linewidth}{@{\extracolsep{\fill}}llllllllr}
\hline
ID & RA & DEC & $m_{I, gal}$ & $m_{I, NSC}$ & n & $R_{e}$ & b/a & PA \\
& (deg) & (deg) & (mag) & (mag) & & (arcsec) &  & (deg) \\
\hline
\multicolumn{9}{c}{Abell 3558}                                                    \\ \hline
dw201987-31475 & 201.987 & -31.475 & 17.80 & 27.63 & 2.12 & 2.73 & 0.87 & 52.56  \\
dw201999-31531 & 201.999 & -31.531 & 18.03 & 24.16 & 1.71 & 1.32 & 0.98 & -57.67 \\
dw201985-31498 & 201.985 & -31.498 & 18.13 & 24.61 & 1.78 & 1.58 & 0.50 & 55.97  \\
dw201989-31499 & 201.989 & -31.499 & 18.25 &       & 1.07 & 1.29 & 0.50 & -89.25 \\
dw202020-31520 & 202.020 & -31.520 & 18.58 & 24.10 & 1.18 & 1.17 & 0.87 & 43.52  \\
dw201876-31372 & 201.876 & -31.372 & 18.64 & 24.29 & 1.39 & 1.18 & 0.87 & -50.65 \\
dw201894-31404 & 201.894 & -31.404 & 18.88 & 25.96 & 0.93 & 2.07 & 0.52 & -87.90 \\
dw201966-31540 & 201.966 & -31.540 & 19.21 & 24.76 & 1.30 & 1.73 & 0.82 & -80.11 \\
dw201868-31375 & 201.868 & -31.375 & 19.25 & 24.17 & 1.53 & 2.38 & 0.88 & 70.74  \\
dw201897-31380 & 201.897 & -31.380 & 19.37 & 24.16 & 1.43 & 1.10 & 0.93 & 4.82   \\
dw201896-31410 & 201.896 & -31.410 & 19.40 & 25.62 & 0.55 & 0.65 & 0.62 & 37.54  \\
dw201995-31498 & 201.995 & -31.498 & 19.42 & 24.45 & 1.70 & 0.94 & 0.93 & -44.51 \\
dw201847-31396 & 201.847 & -31.396 & 19.68 & 25.22 & 0.61 & 1.59 & 0.78 & 72.17  \\
dw202052-31765 & 202.052 & -31.765 & 19.72 & 25.06 & 1.27 & 1.67 & 0.88 & -57.81 \\
dw202001-31546 & 202.001 & -31.546 & 19.76 & 25.22 & 1.16 & 1.31 & 0.91 & -24.19 \\
...            & ...     & ...     & ...   & ...   & ...  & ...  & ...  & ...    \\ \hline
\multicolumn{9}{c}{Abell 1736a}                                                  \\ \hline
dw203690-27132 & 203.690 & -27.132 & 17.98 & 24.83 & 2.88 & 2.79 & 0.72 & -18.03 \\
dw202006-27344 & 202.006 & -27.344 & 18.44 & 23.77 & 1.13 & 1.99 & 0.96 & 87.55  \\
dw201693-27464 & 201.693 & -27.464 & 19.35 & 24.23 & 1.12 & 2.23 & 0.98 & -47.98 \\
dw202039-27353 & 202.039 & -27.353 & 20.13 & 25.56 & 1.00 & 2.22 & 0.48 & -55.97 \\
dw201675-27430 & 201.675 & -27.430 & 20.15 & 23.72 & 0.90 & 0.59 & 0.98 & -42.86 \\
dw201675-27416 & 201.675 & -27.416 & 20.24 & 23.43 & 2.00 & 1.65 & 0.76 & 45.14  \\
dw202039-27357 & 202.039 & -27.357 & 20.50 & 26.72 & 0.90 & 1.18 & 0.74 & 76.33  \\
dw202016-27363 & 202.016 & -27.363 & 20.67 &       & 0.88 & 1.53 & 0.50 & -65.94 \\
dw203718-27119 & 203.718 & -27.119 & 20.72 & 25.00 & 0.75 & 1.70 & 0.92 & -27.68 \\
dw201700-27432 & 201.700 & -27.432 & 20.74 &       & 0.76 & 0.94 & 0.85 & 50.90  \\
dw201709-27471 & 201.709 & -27.471 & 21.03 &       & 0.84 & 1.53 & 0.85 & -28.54 \\
dw203722-27166 & 203.722 & -27.166 & 21.14 & 24.96 & 0.78 & 0.94 & 0.84 & 45.32  \\
dw201698-27468 & 201.698 & -27.468 & 21.23 &       & 1.06 & 1.04 & 0.78 & -52.25 \\
dw201713-27467 & 201.713 & -27.467 & 21.28 & 27.11 & 0.99 & 1.03 & 0.80 & -56.45 \\
dw201682-27417 & 201.682 & -27.417 & 21.40 &       & 0.80 & 0.98 & 0.47 & -84.97 \\
...            & ...     & ...     & ...   & ...   & ...  & ...  & ...  & ...    \\ \hline
\end{tabular*}
\end{table*}

\begin{figure}
    \centering
    \includegraphics[width=\linewidth]{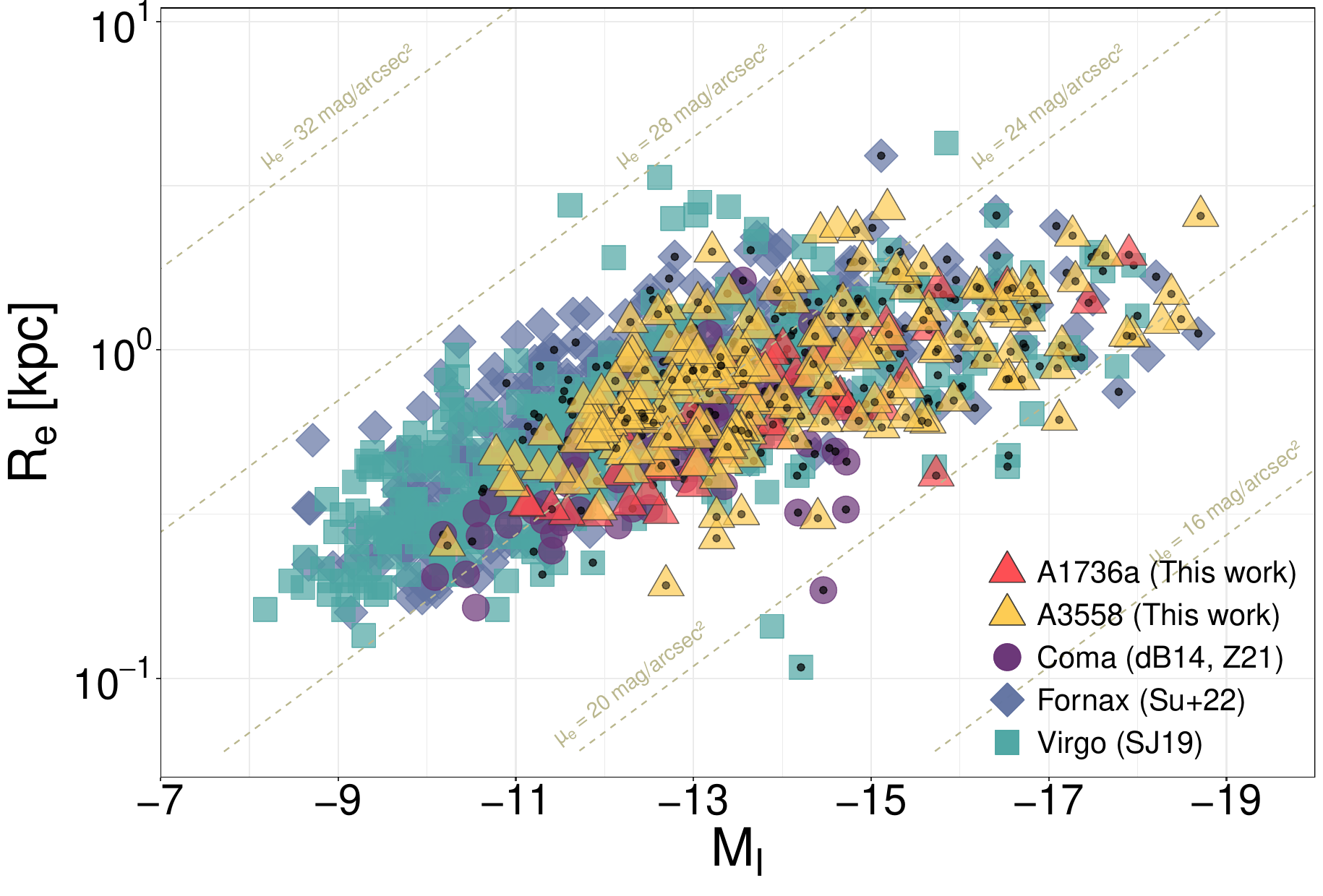}
    \caption{Relation between absolute magnitude and effective radius in the I-band as obtained from \textsc{GALFIT} modelling of the objects selected in the final sample for the A3558 and A1736a clusters (yellow and red triangles, respectively). For reference, also shown are the Coma galaxy sample from \citetalias{zanatta21} (which includes the sample from \citetalias{denbrok14}, purple circles), the Fornax sample from \citet{su22} (navy blue diamonds) and the Virgo sample from \citetalias{sanchez-janssen19b} (cyan squares). Nucleated objects are represented by markers with solid black circles at the centre. The objects detected in the SSC in this work follow the scaling relation of size vs luminosity observed for dwarf galaxies in other galaxy clusters.}
    \label{fig:magxre}
\end{figure}

\subsection{Galaxy and NSC modelling}
\label{sec:modelling}

To determine the structural and photometric properties of the candidate galaxies and their NSCs, we model their surface brightness profiles using \textsc{galfit} \citep{peng02}, employing the same approach used to model faint Coma galaxies in \citetalias{zanatta21}.  

We use the segmentation maps from the first \textsc{SourceExtractor} run to mask all objects around our detected galaxies, except for the central point sources in the visually identified nucleated dwarfs. We also use PSF models obtained with \textsc{PSFex} \citep{bertin11}. Initial conditions for the fits are based on \textsc{mag\_auto} magnitudes and \textsc{flux\_radius} results from the \textsc{SourceExtractor} catalogue, in addition to an initial S\'ersic index of $n = 0.75$, position angle of 45 degrees and axis ratio of 0.8. Non-nucleated galaxies are modelled with a single S\'ersic profile, while nucleated ones are modelled with a S\'ersic profile with the addition of a PSF component to model the NSC. We model nuclei as point-sources based on the fact that NSCs have typical sizes of 1-50 pc \citep{sanchez-janssen19, neumayer20}, which given the pixel scale of HST/ACS translates in all but the largest NSCs to be unresolved in our images. In some rare cases we find that visually identified nucleated galaxies are better modelled by two S\'ersic profiles, instead of a Sérsic plus PSF component. In this case we adopt as the galaxy structural and photometric parameters the ones from the S\'ersic Profile with largest estimated effective radius. In this setup, it is to be expected for some of the galaxy light to contribute to the estimated NSC magnitude, since the two S\'ersic profiles overlap each other. However, we have found galaxies fitted this way to show no significant discrepancies in their final \textsc{GALFIT} estimated structural parameters from the general trends observed for nucleated galaxies (see the lack of strong outliers in Figures \ref{fig:magxre} and \ref{fig:nucl}), and this affects less than three objects in A3558. Furthermore, $HST$ studies of NSCs in nearby clusters show that stellar nuclei are rarely offset from the geometric centre of the host galaxy \citep{cote03,turner12}, therefore to aid in the modelling of fainter objects we constrain the relative position of the components within three pixels of each other. This is also done to guarantee an insignificant probability of contamination from projected GCs over the galaxy centre. We estimate the contamination by counting the total number of point sources (\textsc{class\_star} $ \geq 0.6$) that have a magnitude difference of less than 0.5 mag with respect to each NSC. We them multiply the surface density of these candidates by the area enclosed in a circle of three pixels in radius. This results in the mean number of contaminants to be $\approx 0.002$ per galaxy. 

Moreover, 40 galaxies within our sample for A3558 and 14 in A1736a are found to have more complex structures by the fact that the \textsc{galfit} residuals present significant structures, such as embedded disks, bulges and twisted isophotes. To reach a good \textsc{galfit} model for such objects would lead to convoluted and over-complicated modelling beyond the scope of this work. Nucleation classification in such objects via visual inspection is also very difficult, as bulges and embedded disks can very easily lead to false NSC detections. A simple aperture photometry procedure indicates that all galaxies not well fitted with \textsc{galfit} are brighter than $M_I < -19.0$ mag. With the exception of the H21 sample, all data of the other environments from the literature used in this work employ magnitude cuts at the bright-end at similar luminosities and for similar reasons (\citetalias{denbrok14}; \citealt{munoz15}; \citetalias{sanchez-janssen19b};  \citealt{carlsten20b, carlsten22a}). Therefore, due to the aforementioned modelling and nuclear classification difficulties and to keep our comparison homogeneous between all environments, we consider in this work only galaxies in between $-7.5 > M_I > -19.0$. This affects the bright-end of A3558, A1736a and the faint and bright end of H21. The reason for the faint cut in the H21 sample is due to the behaviour of the logistic relation, which will be discussed in the next section. Nevertheless, after magnitude cuts, we are left with 180 objects in A3558 and 44 in A1736a, preserving 85\% and 75\% of the detected galaxies in the previous step, for each environment, respectively. Of these, 104 are found to be nucleated in A3558 and the same is true for 16 objects in A1736a. In Table \ref{tab:catalog}, we show the final catalogue for the A3558 and A1736a galaxy sample used in the rest of our analysis, as well as several photometric properties from the \textsc{galfit} modelling. Magnitudes are dereddened from Galactic extinction using the maps from \citet{schlafly11} and a summary of the method is shown in Figure \ref{fig:detectionpanel}. 

\begin{figure}
    \centering
    \includegraphics[width=\linewidth]{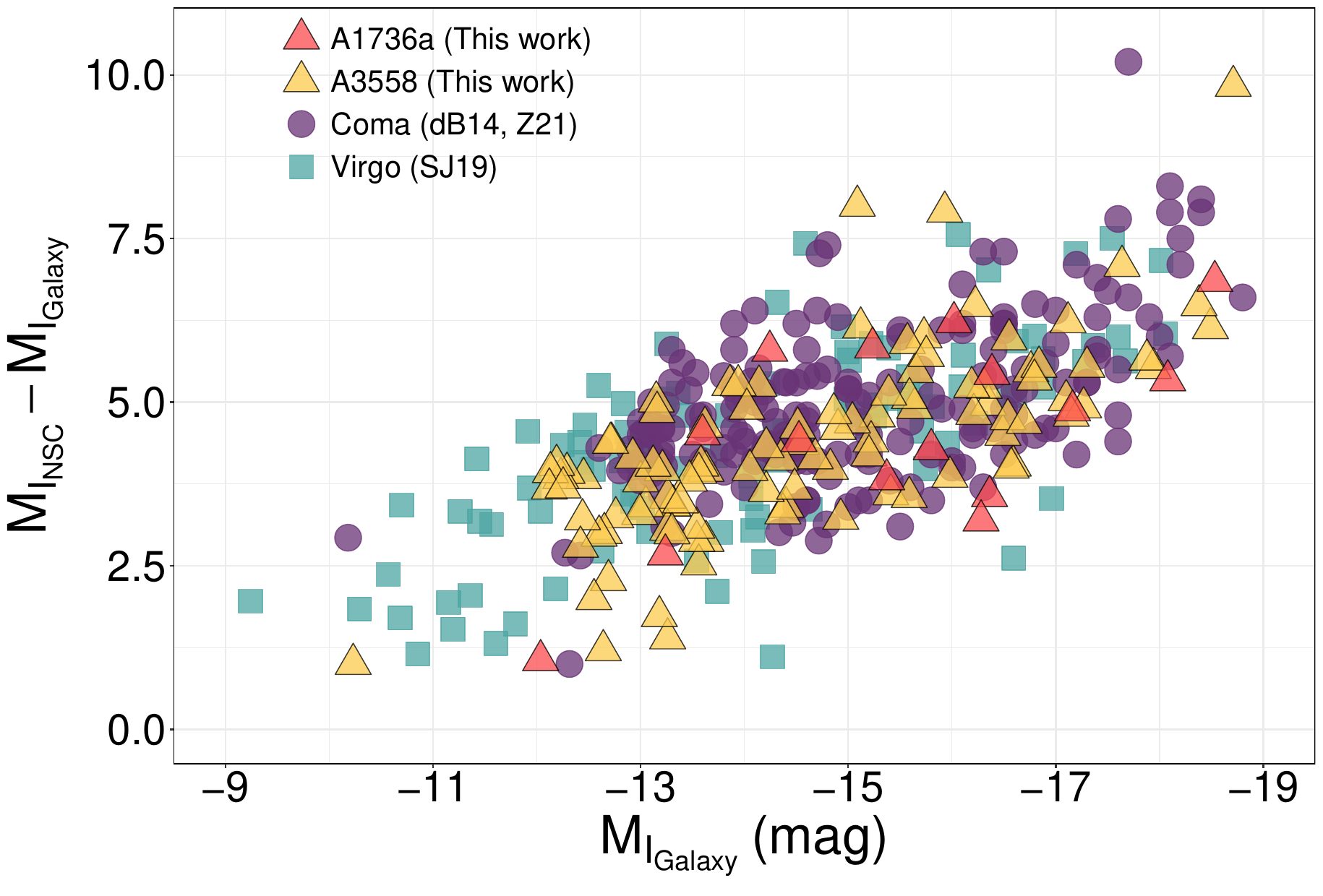}
    \caption{Difference between the magnitude of the nuclei, $M_{I, NSC}$ and the magnitude of its host galaxy, $M_{I, Galaxy}$, for all nucleated galaxies in the A3558 and A1736a clusters sample from this work (shown in table \ref{tab:catalog}) as well as nucleated galaxies from the Coma cluster samples of \citetalias{zanatta21} and \citetalias{denbrok14} and nucleated galaxies in the Virgo cluster (from \citetalias{sanchez-janssen19b}), as a function of the host galaxy absolute magnitude. Colours and markers are the same as Figure \ref{fig:magxre}. For all environments, brighter galaxies tend to show larger differences in magnitude from their nuclei, although a scatter is also evident, showcasing its stochastic nature.}   
    \label{fig:nucgal}
\end{figure}

Despite careful visual inspection, there is still the possibility of contaminants being present in our galaxy samples. The spherical morphology of background early-type galaxies at specific distances can lead to objects to be mistakenly identified as a dwarf spheroidals at SSC distance, and background disk galaxies may appear visually similar to compact low surface brightness objects. In Figure \ref{fig:magxre}, we show the relation between the absolute magnitude and effective radius in the I-band modelled with \textsc{galfit} for the final A3558 and A1736a samples detected in this work. We show as well objects in the Coma, Fornax and Virgo samples, for reference. The loci of our SSC detections in this graph indicate a compatible magnitude-size relation to objects in the literature samples. Therefore, we conclude that our SSC catalogues comprise only galaxies not likely to be contaminant background objects. 

In Figure \ref{fig:nucgal}, we show the difference between the magnitude of the NSC and its host galaxy for all the nucleated galaxies. We show this difference for galaxies in A3558 and A1736a, as well as Coma and Virgo. The relative contribution of the NSC to the overall brightness of the host galaxy decreases with galaxy luminosity, albeit with a large scatter in the relation. From the perspective of comparing different environments, we see that the trend in Figure \ref{fig:nucgal} is very similar for all the clusters. In \citetalias{sanchez-janssen19b} and \citetalias{zanatta21}, it is argued that such results indicate that the luminosity of the NSC is related to that of the host galaxy regardless of the environment, but the large scatter in the relation indicates that NSC growth varies substantially from one galaxy to another. We find this conclusion consistent with the NSC data from the SSC clusters analysed in this work. 

It is also relevant to note that galaxy detections in A3358 and A1736a drop significantly at fainter magnitudes than $M_I = -11.0$ mag. In both Figure \ref{fig:magxre} and Figure \ref{fig:nucgal} this can be observed but not to be the case for the Virgo and Fornax samples. A relative decrease in completeness in the detection of faint dwarfs in the SSC is to be expected given that the Virgo and Fornax galaxy clusters are considerably closer. Nevertheless, in what concerns the main objective of this work, i.e., to compare nucleation statistics across different environments, we do not expect this drop in completeness at the lower end of magnitude to be relevant since the likelihood of galaxy nucleation have been observed to universally become almost negligible for magnitudes fainter than $M_I \sim -10.5$ mag (\citetalias{sanchez-janssen19b}; \citet{neumayer20}; \citetalias{zanatta21}).  

\section{Statistical Modelling of the nucleation fraction}
\label{sec:logit}

To analyse the nucleation fraction in the A3558 and the A1736a clusters and how it compares to other environments, we employ Bayesian logistic regression \citep[see, e.g.][for a detailed description]{Hilbe2017}. In \citetalias{zanatta21}, we employed this statistical approach successfully in order to avoid arbitrary binning of the data. This is essential to properly study how the nucleation fraction varies over the entire range of magnitudes and environments. By estimating such quantity in a homogeneous way for all environments with a robust and adequate methodology, we produce smooth, continuous and realistic models based on the observational data. This provides a clear panorama of the matter and enables a direct and honest comparison between different datasets. On the other hand, an approach based on the so-called "best-fit" linear model of arbitrarily binned data points is always prone to small completeness variations along the range of the predictor variable producing misleading local non-linearities that can potentially bias an entire analysis. 

Logistic regression belongs to the family of generalised linear models, and is particularly suitable for handling Bernoulli-distributed (binary) data. Such distribution characterises processes with two possible outcomes $\{0,1\}$, be it success or failure, yes or no, or alike -- in our case, it is nucleation or non-nucleation. Apart from the application of logistic regression to model the nucleation fraction in \citetalias{zanatta21}, previous applications of logistic models in Astronomy include the studies of star-formation activity in primordial dark matter halos \citep{deSouza2015AeC}, the escape of ionising radiation at high-redshift \citep{Hattab2019}, the effect of environment in the prevalence of Seyfert galaxies \citep{desouza16}, the redshift evolution of UV upturn galaxies \citep{Dantas2020} and the GC occupation fraction in dwarf galaxies \citep{eadie22}.

The full behaviour of the nucleation fraction is significantly distinct from the logistic relation \citepalias{zanatta21}, with a peak estimated at masses log($\mathcal{M}/\mathcal{M}_{\odot}$) $\approx$ 9 (\citetalias{sanchez-janssen19b}; \citealt{neumayer20}) declining toward higher and lower masses. To mitigate this issue, in addition to the magnitude cuts at the bright end ($M_I > -19.0$ mag) explained in the previous section, we also remove galaxies fainter than $M_I = -7.5$ mag. At such and fainter magnitudes, the nucleation fraction is expected to be insignificantly small and virtually constant, especially at the Local Volume (\citealt{ordenes-briceno18}; \citetalias{sanchez-janssen19b, zanatta21}, \citealt{hoyer21}). The magnitude cut at the faint-end only affects the H21 sample. With these constraints, we ensure that for all studied environments, the data present a homogeneous magnitude range, with reliable photometry and nucleation classification as well as being within the range of galaxy luminosity/mass where the nucleation fraction is universally monotonically crescent \citepalias{sanchez-janssen19b, zanatta21}, and therefore well-suited to be modelled with a logistic approach.     

The regression model is the following:
\begin{align}
\label{eq:model}
   &\rm y_{i}\sim \rm {Bern}\left(p_{i}\right), \notag   \\
   &\eta_{i} \equiv \log\left(\frac{p_{i}}{1-p_{i}}\right),\\
    & \rm \eta_{i} =  \beta_{1[k]}+\beta_{2[k]}\,M_{I,i},\notag \\
   & \begin{bmatrix}\notag
\beta_{1[k]}     \\
\beta_{2[k]}     
    \end{bmatrix}
    \sim \rm {Norm}{\left(   \begin{bmatrix}\notag
\mu_{\beta}     \\
\mu_{\beta}     
    \end{bmatrix},\Sigma\right); \quad \Sigma \equiv 
    \begin{bmatrix}\notag
\sigma_{\beta}^2   & 0  \\
0  & \sigma_{\beta}^2  
    \end{bmatrix}},\\
    &\mu_{\beta} \sim \rm{Norm(0,10^2)}; \quad \notag
   \sigma_{\beta}^2 \sim \rm{Gamma(0.1,0.1)}.
    \end{align}
The above model reads as follows. Each of the $i$-th galaxies in the dataset has its probability of manifesting nucleation, $y_i$, modelled as a Bernoulli process, whose probability of success relates to its absolute magnitude in the I-band, $M_{I,i}$, through a logit link function, $\eta_{i}$ (to ensure the probabilities will fall between 0 and 1),  where the index $k$ encodes the cluster/group environment. We adopt multi-Normal priors for the intercept $\beta_{1[k]}$ and slope $\beta_{2[k]}$ coefficients with a common mean $\mu_{\beta}$ and variance $\sigma_{\beta}^2$, to which we assigned weakly informative Normal and Gamma hyper-priors respectively. The latter is important in order to implement a hierarchical partial pooling framework. In such methodology, different parameters are inferred for each environment but are allowed to share information. We refer the reader to Sec. 4 of \citetalias{zanatta21} for a brief discussion of the advantages of this choice and \citet{Hilbe2017} for details. 

We evaluate the model using the Just Another Gibbs Sampler (\textsc{jags}\footnote{http://cran.r-project.org/package=rjags}) package within the \textsc{R} language \citep{rcore19}. We initiate three Markov Chains by starting the Gibbs sampler at different initial values sampled from a Normal distribution with zero mean and standard deviation of 100. Initial burn-in phases were set to 5,000 steps followed by 20,000 integration steps, which are sufficient to guarantee the convergence of each chain, following Gelman-Rubin statistics \citep{gelman-rubin92}.

\begin{table*}
\centering
\caption{Summary of the parameters estimated from the model presented in Eq. \ref{eq:model}. In the first column, we show $M_{I,f_{n50}}$, the magnitude at which the estimated probability of nucleation reaches 50\%. In the second column, $\Delta$Odds represents the expected change in the odds of nucleation by a variation of one unit of magnitude. In the last two columns, $\beta_1$ and $\beta_2$ are the mean posteriors for the intercept and slope, respectively, of the link function $\eta_{i}$.}

\label{tab:model}
\begin{tabular*}{\linewidth}{@{\extracolsep{\fill}}rrcrr}
\hline
 & $M_{I,f_{n50}}$ & $\Delta$Odds (\%) & $\beta_{1}$ (intercept) & $\beta_{2}$ (slope)\\\\
 \hline
\multicolumn{1}{l}{A3558}              & $-13.17${\raisebox{0.5ex}{\small$_{0.52}^{0.61}$}} & -46.80 & $ -8.31 \pm 1.57$ & $-0.63 \pm 0.116$\\
\multicolumn{1}{l}{A1736a}             & $-15.01${\raisebox{0.5ex}{\small$_{2.30}^{1.06}$}} & -46.84 & $ -9.48 \pm 2.95$ & $-0.63 \pm 0.209$\\
\multicolumn{1}{l}{Coma}               & $-13.00${\raisebox{0.5ex}{\small$_{0.43}^{0.59}$}} & -52.37 & $ -9.64 \pm 1.54$ & $-0.74 \pm 0.110$\\
\multicolumn{1}{l}{Virgo}              & $-14.37${\raisebox{0.5ex}{\small$_{0.58}^{0.46}$}} & -47.50 & $ -9.26 \pm 0.97$ & $-0.64 \pm 0.073$\\
\multicolumn{1}{l}{Fornax}             & $-13.98${\raisebox{0.5ex}{\small$_{0.45}^{0.36}$}} & -51.09 & $-10.00 \pm 0.99$ & $-0.72 \pm 0.076$\\
\multicolumn{1}{l}{Local Volume (H21)} & $-14.83${\raisebox{0.5ex}{\small$_{1.34}^{0.95}$}} & -39.90 & $ -7.55 \pm 1.28$ & $-0.51 \pm 0.095$\\
\multicolumn{1}{l}{Local Volume (C22, Early-Type Hosts)} & $-14.67${\raisebox{0.5ex}{\small$_{2.64}^{1.16}$}} & -35.77 & $ -6.49 \pm 1.51$ & $-0.44 \pm 0.119$\\
\multicolumn{1}{l}{Local Volume (C22, Late-Type Hosts)}  & $-15.47${\raisebox{0.5ex}{\small$_{2.08}^{1.02}$}} & -48.96 & $-10.40 \pm 2.03$ & $-0.67 \pm 0.153$\\
\hline
\end{tabular*}
\end{table*}

\section{Results}
\label{sec:results}

\begin{figure}
    \centering
    \includegraphics[width=\linewidth]{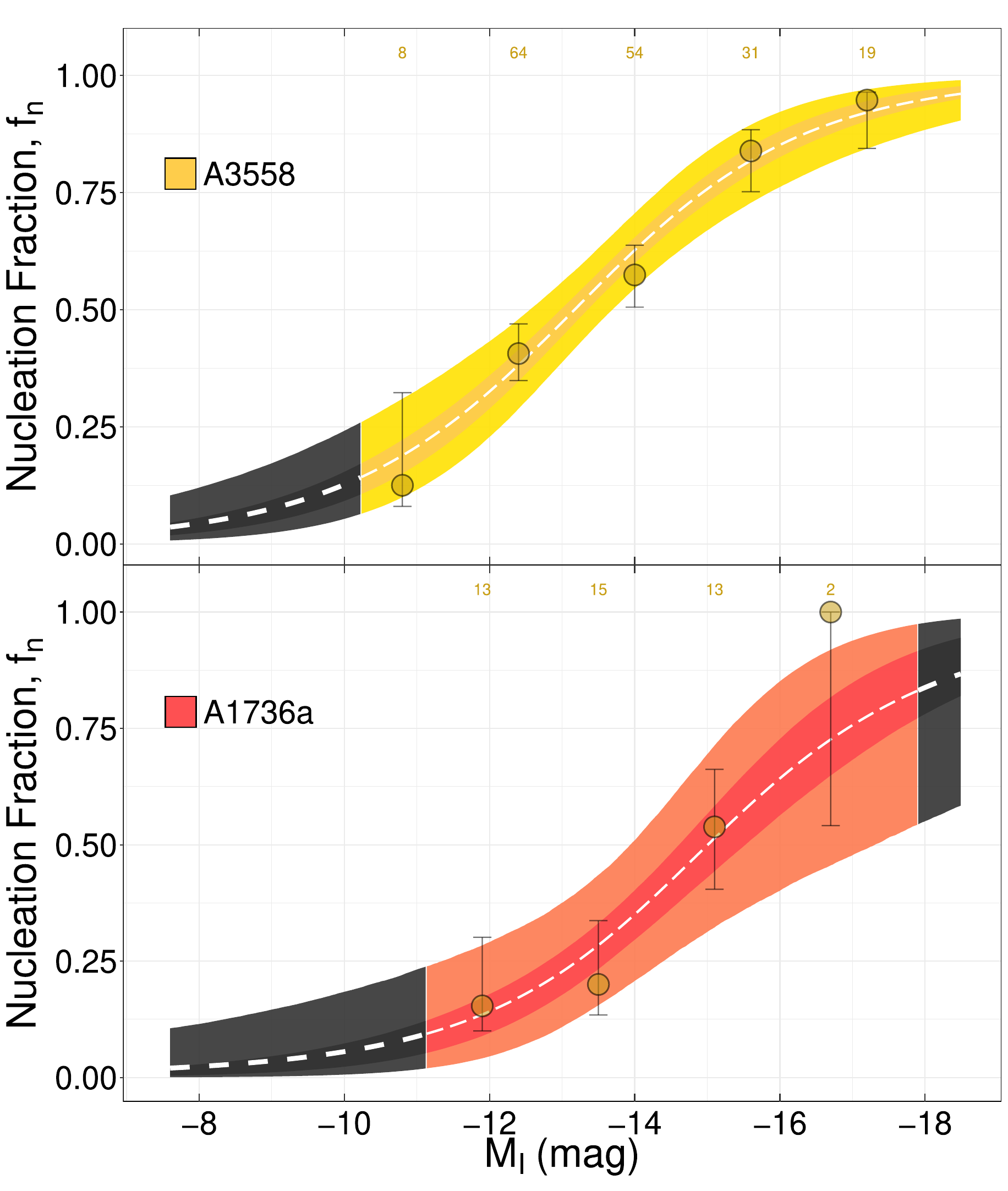}
    \caption{Nucleation fraction versus absolute magnitude for galaxies in A3558 (top) and A1736a (bottom). The white curve represents our estimation of the nucleation fraction based on the Bayesian logistic regression. The coloured shaded regions show the 50\% and 95\% confidence intervals, whereas the grey shades indicate the magnitudes where the model extrapolates the data. The orange solid circles represent the median nucleation fraction in a binned representation of the data, with uncertainties given by the corresponding 68\% Bayesian credible intervals. The number of objects in each bin is shown at the bottom.}   \label{fig:nucl}
\end{figure}

\begin{figure*}
    \centering
    \includegraphics[width=\linewidth]{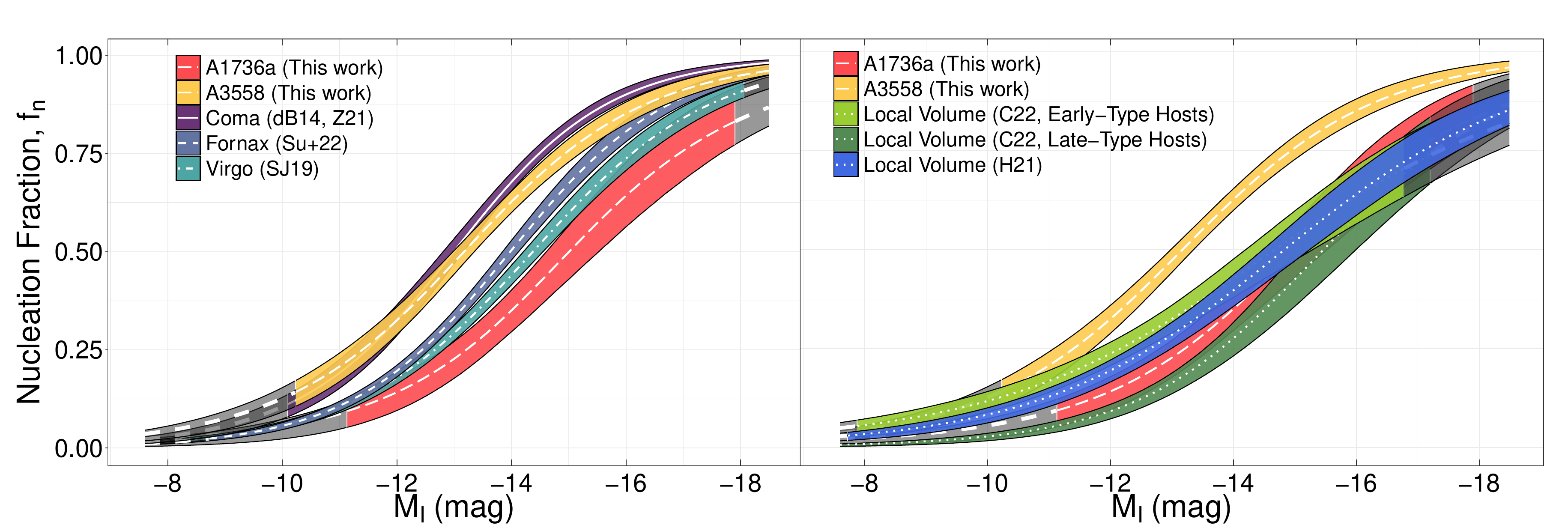}
    \caption{Nucleation fraction as a function of absolute magnitude in the I-band estimated for different environments via Bayesian logistic regression. Coloured filled areas represent the 50\% confidence intervals. In both panels, we show the nucleation fraction for the galaxies in the Shapley clusters presented in this work, A3558 and A1738a, shown in Table \ref{tab:catalog}. On the right panel, we show only galaxy \textit{clusters} compared to A3558 and A1736, while at the left panel, we show the Shapley clusters compared to the Local Volume samples, which comprise galaxies in \textit{groups} and the \textit{field}. Complete references for the literature data are shown in Table \ref{tab:appendixA}. A3558 galaxies show a nucleation fraction as high as Coma, especially at the fainter end. On the other hand, results for A1736a indicate a nucleation fraction slightly lower than galaxies in Virgo and Fornax but compatible with the $f_{n}$ observed in galaxies at the Local Volume. Overall, except for the Local Volume, the rate at which the probability of nucleation changes with magnitude is similar for all environments, while the probability of nucleation at fixed magnitude shows a clear dependence on the environment, being highest in the more massive environments, A3558 and Coma.}

    \label{fig:nuclboth}
\end{figure*}

In this section, we present the nucleation fraction modelled for the A3558 and A1736a datasets using the logistic regression methodology described in the previous section. In order to compare the nucleation in the SSC to other environments, we also present the nucleation fraction for Coma, Virgo and Fornax galaxy clusters, as well as the nucleation fraction for the Local Volume samples C22 and H21. The main results of this section are shown in Figure \ref{fig:nuclboth}. 

We note that the results presented in this section are a direct follow-up to the ones presented in \citetalias{zanatta21}. In summary, they presented the high nucleation fraction of Coma cluster galaxies when compared to galaxies in Virgo and Fornax clusters as well as to Local Volume samples. These results indicate a clear dependence between environment and nucleation probability. This is evidenced by the fact that, in \citetalias{zanatta21}, while the rate that nucleation varies with luminosity has been found to be the same in all cases, galaxies in more massive environments were found to be more likely to host NSCs for a fixed luminosity/mass. In this section, we show these findings are confirmed with the inclusion of nucleation information for dwarf galaxies at the A3558 and A1736a clusters in the SSC.

\subsection{Nucleation fraction in the Abell 3558 and Abell 1736a clusters in the Shapley Supercluster}

In Figure \ref{fig:nucl}, we show the estimated nucleation fraction, $f_n$, as a function of galaxy absolute magnitude, $M_{I}$, for the A3558 cluster (yellow curve, top panel) and the A1736a cluster (red curve, bottom panel). The mean of the estimated model parameters and associated 68\% confidence intervals are shown in Table \ref{tab:model}. Additionally, we show a binned representation of the observed data as orange-coloured markers with associated uncertainties from the 68\% Bayesian credible intervals. Immediately it is clear that the logistic regression is able to capture the behaviour of the nucleation fraction in the entire magnitude range observed and provides a much clearer visualisation of the matter due to its presentation as a smooth solution. Furthermore, grey regions indicate magnitudes where the model is able to extrapolate the data in order to achieve probabilities between 0 and 1. In agreement with the conclusions of previous studies (\citetalias{sanchez-janssen19b}; \citealt{neumayer20}; \citetalias{zanatta21}), we find a strong dependence of the nucleation fraction on galaxy luminosity. In both SSC clusters, it peaks at nearly $f_n \approx$ 100\% at $M_{I} \approx -18$ mag (the characteristic luminosities of classical dwarf elliptical galaxies) and then declines to become almost negligible at $M_{I} \approx -10$ mag. In A3558, the nucleation fraction is noticeably high, with more than half of the $M_{I} = -13$ mag galaxies still hosting NSCs. For the A1736a nucleation fraction, our results indicate a steeper increase in nucleation with magnitude, although the smaller sample size at the bright end results in larger uncertainties. Overall, at fixed galaxy magnitudes and within uncertainties, the nucleation fraction in A1736a is systematically lower than in A3558. To reach a 50\% nucleation fraction probability, galaxies in A1736a need to be almost two magnitudes brighter than in A3558. 

\subsection{Comparison with Other Environments}

In Table \ref{tab:model}, we show the final estimated parameters from the logistic regression for all environments and two indexes useful to interpret the results. The first is the half-nucleation magnitude, $M_{i, f_{n50}}$, which is the magnitude at which there is an estimated 50\% probability for a galaxy to be nucleated in a given environment. The second is the $\Delta$Odds index, which represents the expected change in the nucleation probability by a variation of one magnitude unit (see \citet{desouza16} for additional details). Then, in Figure \ref{fig:nuclboth}, we show the mean posterior for the nucleation fraction in the A3558 and A1736a clusters in the SSC compared to that in other galaxy clusters analysed in this work (left panel), as well as in the Local Volume (right panel). 

The C22 sample is divided and stacked based on the morphology of the central galaxy of groups to which objects in the sample are associated to. The H21 sample cannot be reliably divided in the same fashion for the difficulty in assigning hosts to this sample galaxies, as explained in Section \ref{sec:data}.

The nucleation fraction for Coma, Virgo and Fornax environments have already been analysed, with the same methodology, in \citetalias{zanatta21}. Small variations in the estimated parameters in that work to the ones shown here are to be expected due to the employment of partial pooling and the inclusion of new environments in the model. However, there is no significant difference between the logistic results for these environments between this work and \citetalias{zanatta21}, with all estimated parameters for Coma, Virgo and Fornax being equal, within uncertainties, between both works.  

From the comparison of the nucleation in A3558 and A1736a with all the other environments studied in this work, two main results can be noticed: 

\begin{itemize}
    \setlength\itemsep{1em}

    \item[1)] Within uncertainties, the nucleation fraction in the central cluster of the SSC, A3558, shows a remarkable similarity to the nucleation fraction for galaxies in the Coma cluster and is significantly higher than every other environment.  

    \item[2)] Conversely, the estimated nucleation fraction for the A1736a cluster is significantly lower than in A3558 (and Coma). At the bright end ($M_{I} \gtrsim -15$), within uncertainties, the probability of nucleation in A1736a galaxies can be compared to that of the Virgo and Fornax clusters, albeit systematically lower. At the right panel of Figure \ref{fig:nuclboth}, we can see, however, that overall the estimated nucleation fraction for A1736a is at a very similar loci to the ones estimated for Local Volume galaxy samples.      

\end{itemize}

Furthermore, in what regards to the general panorama of galaxy nucleation across the different environments analysed: 

\begin{itemize}

    \item[1)] In the left panel of Figure \ref{fig:nuclboth}, we can see that for all galaxy \textit{clusters}, nucleation has a very similar dependence on dwarf galaxy luminosity. Nucleation fraction curves for these environments present compatible slopes, with $\beta_{2}$ parameters consistent with each other at the 1\,$\sigma$ level. The $\Delta$odds indexes, shown in Table \ref{tab:model}, are also within $\sim$ 5\% of difference between A3558, A1736a, Coma, Virgo and Fornax.

    \item[2)] At the right panel of Figure \ref{fig:nuclboth} one can see that the estimated nucleation fraction for Local Volume samples present a slightly different slope when compared to galaxy clusters. The C22 sample comprising satellites of Early-Type hosts and the H21 sample present a $\Delta$odds difference of less than 3\% in between themselves, but depart $\sim$ 9\% and $\sim$ 13\% from the mean $\Delta$odds of galaxy clusters, respectively. On the other hand, the nucleation fraction for C22 sample of dwarfs around Late-type hosts presents a slope compatible with that of galaxy clusters.  

\end{itemize}    

Our results indicate a clear distinction between the nucleation fraction for dwarfs satellites of early and late-type hosts in the C22 sample. Objects associated with Late-type hosts present the lowest nucleation fraction of all environments analysed, but with a comparable $\Delta$Odds index to galaxy clusters. The nucleation fraction for C22 galaxies around Early-type hosts, however, display $\Delta$Odds estimates only comparable to the H21 Local Volume sample. In fact, for both C22 galaxies around early-type hosts and the H21 sample, galaxies fainter than $M_I \approx -15$ mag present an estimated nucleation fraction comparable to clusters such as Virgo and Fornax. At $M_I \approx -11$, Local Volume H21 and C22 galaxies around early-type hosts present an estimated nucleation fraction comparable to even galaxies in A3558 and Coma, around a 10\%, while such quantity is already negligible for C22 galaxies around late-type hosts. We discuss these results in more detail in the next section and in Appendix \ref{appendixC}. In the latter we show that the nucleation fraction of the entire C22 sample, i.e., not divided between early and late-type hosts, is remarkably similar to same quantity estimated for the H21 sample.    

\subsection{The Role of Environment in the Galaxy Nucleation Fraction}

\begin{figure}
    \centering
    \includegraphics[width=\linewidth]{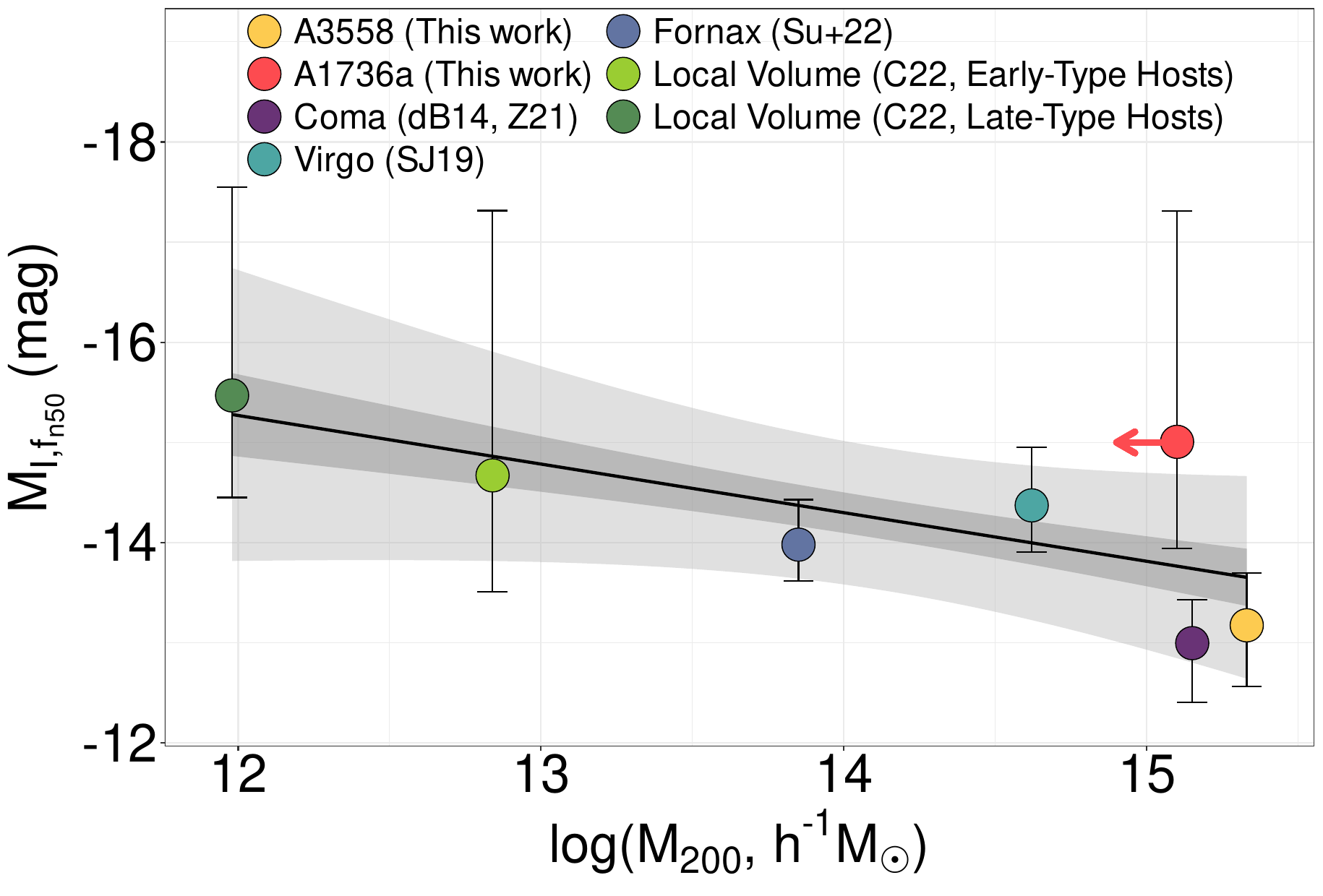}
    \caption{Estimated magnitude at which the probability of nucleation is 50\%, $M_{I,fn50}$, as a function of host virial halo mass (references mentioned in the text). Colours are the same as in Figure \ref{fig:nuclboth}. The mean, 50\% and 95\% confidence intervals for the linear model are indicated by the solid line and grey-shaded areas, respectively. The vertical error bars indicate uncertainties from the logistical analysis. More massive environments exhibit fainter values of $M_{I,fn50}$, especially if we consider that the halo mass estimate for A1736a is very likely overestimated, as mentioned in the text, and represented in the figure by the red arrow.}
    \label{fig:hostmass}
\end{figure}

In Figure \ref{fig:hostmass} we present a linear regression based on the relation $M_{I,fn50} = \alpha + \beta\, \text{log}(\mathcal{M}_{200})$, where $\mathcal{M}_{200}$ are halo mass estimates from the literature (detailed in Section \ref{sec:data}). The H21 sample is not included in this figure as the catalogue from \citet{hoyer21} does not objectively present host galaxies associated with the detected objects and includes a significant portion of isolated field dwarfs (see Appendix \ref{appendixC} for details). Therefore we are not able to assign an estimated halo mass for this sample. Another important caveat is that in the case of A1736a, due to the lack of a specific A1736a halo mass estimate in the literature, we adopt the halo mass estimate for the entire Abell 1736 cluster (i.e., A1736a \textit{and} A1736b) from \citet{lopes18}. Hence, our adopted A1736a halo mass estimate is an obvious overestimation. This is represented by a leftwards red arrow associated with the A1736a marker in Figure \ref{fig:hostmass}. 

As observed in \citetalias{zanatta21}, early-type dwarfs in the Local Volume show the highest values of $M_{I,f_{n,50}}$. Dwarfs around Late-type galaxies in the C22 sample are found to display a 50\% chance of being nucleated as bright as $M_I \approx$ -15.5 mag, while in the most massive environments analysed, such as Coma, $M_{I,f_{n,50}}$ is as faint as $M_I \approx$ -13 mag. A3558 has an estimated halo mass compatible, within uncertainties, with that of the Coma cluster (log($\mathcal{M/M\odot}$) $\approx 15.33$ for A3558 \citep{lopes18} and $\approx 15.10$ for Coma \citep{lokas03}). Remarkably, this translates into a very similar estimated $M_{I,f_{n,50}}$ between these two environments: the two faintest across all environments where galaxy nucleation have been studied to date. In the case of A1736a, its estimated $M_{I,f_{n,50}}$ is brighter than that estimated for the C22 dwarfs satellites of Early-Type galaxies and slightly fainter than that of dwarf satellites around Late-Type galaxies. 

The proposed linear model is able to efficiently fit the relation between $\mathcal{M}_{200}$ and $M_{I,f_{n,50}}$. Within uncertainties, all data points are within the model 50\% confidence intervals. The result confirms the findings from \citetalias{zanatta21}, i.e., that \emph{nucleation is more common in more massive haloes regardless of dwarf galaxy luminosity}. In fact, A3558 has a value of $M_{I,f_{n50}}$ that differs by a factor of at least 10 in luminosity to Local Volume groups while also presenting the highest estimated halo mass among the environments analysed in this work. Considering the aforementioned known overestimation of the A1736a halo mass, its loci in Figure \ref{fig:hostmass} is compatible with the estimated linear relation within uncertainties, however, if we adopt this estimated $\mathcal{M}_{200}$ and $M_{I,f_{n,50}}$ relation, we can infer that a more precise halo mass estimate for A1736a should be similar to that adopted for the dwarf satellites around Early-Type galaxies in the C22 sample. 

\section{Discussion}
\label{sec:discussion}

In this work we used a new catalogue of dwarf galaxies in the SSC as well as literature data for dwarfs in the Coma, Virgo and Fornax clusters and in the Local Volume to show that the fraction of nucleated dwarfs at a fixed luminosity depends substantially on the environment. This confirms the results presented in \citetalias{zanatta21}. The extremely dense and massive environment at the centre of the SSC, A3558, presents a decisively high nucleation fraction over the entire range of galaxy luminosity considered in this work. The only other environment as massive as A3558 outside of the SSC is the Coma cluster, which presents a nucleation fraction remarkably similar to A3558. Moreover, the less massive A1736a cluster expectedly presents the lowest nucleation fraction of all galaxy clusters considered, but is comparable with the $f_n$ estimated for Local Volume groups. In this section we attempt to contextualise these results within the current literature discussion regarding the nucleation of dwarf galaxies. 

\subsection{Dependence of $f_{n}$ on mass/luminosity}
\label{sec:discussion-lumi}

It was observed in environments such as Virgo \citepalias{sanchez-janssen19b}, Fornax \citep{ordenes-briceno18, su22} and the Local Volume \citep{hoyer21}, that the galaxy nucleation fraction displays a peak for masses log($\mathcal{M/M\odot}$) $\approx 9$, approximately $M_I \approx -18$, then declining for lower and higher masses/luminosities. In this study, we model the nucleation fraction of quiescent dwarf galaxies within $-7.5 < M_I < -19.0$, i.e., the approximate luminosity range where the nucleation fraction is known to be monotonically crescent, at least for early-type dwarfs \citep{neumayer20}. This enables us to reliably model the observed nucleation fraction as a smooth logistic fit, in order to compare this quantity across different environments without resorting to any kind of arbitrary binning. Using this methodology, in Figures \ref{fig:nucl} and \ref{fig:nuclboth} it becomes immediately clear the nearly-universal dependence of the nucleation fraction with galaxy luminosity. In \citetalias{zanatta21} this dependence was found to be remarkably homogeneous across the Coma, Virgo, Fornax and Local Volume samples. The present work results, with updated Fornax and Local Volume samples as well as the new data from two SSC clusters, A3558 and A1736a, confirms this universal aspect. 

In other words, this dependence implies that it is more likely for more massive dwarfs to accumulate material at their central parsecs in the form of dense bound clusters. The NSC formation and growth via star cluster inspiral seems to be the dominant process for low-mass quiescent galaxies in the luminosity/mass range studied in this work \citep{neumayer20, fahrion22, su22}. This is evidenced by the fact that NSCs in faint, early-type dwarfs tend to be more metal-poor than their host galaxies \citep{fahrion20, johnston20}, their GC and NSC occupation fraction is remarkably similar \citepalias{sanchez-janssen19b} and models of GC inspiral in this luminosity regime predict, with excellent agreement, the observed scaling of NSC mass with galaxy mass (\citealt{antonini13, gnedin14}; \citetalias{sanchez-janssen19b}; \citealt{leaman21, fahrion22}). To be able to form NSCs, the GC inspiral scenario requires GCs massive enough to survive tidal dissolution as a consequence of the dynamical friction that ensues their orbital decay \citep{tremaine75, capuzzo-dolcetta93, bekki04, antonini13}. Therefore, NSC growth is intrinsically related to the GC mass function (GCMF). Furthermore, this process also needs to happen in less than a Hubble time, so factors such as GC formation distance and the galaxy structural properties also play an important role \citep{arca-sedda14,  neumayer20, leaman21}. In summary, to explain the observed trend with galaxy luminosity the GCMF of fainter dwarfs must be relatively bottom-heavier or such galaxies form GCs preferentially distant from the galaxy centre. The latter could be explained if low-mass dwarfs would be, on average, more diffuse and extended, but this contradicts the observed mass-size relation of faint dwarfs in the nearby Universe \citep{mcconnachie12, eigenthaler18, ferrarese20}. 

In this context, an important aspect of the methodology applied in this work is the fact that it enables the calculation of indexes such as $M_{I,f_{n,50}}$ and $\Delta$Odds in order to easily and precisely compare the nucleation in different environments. The $\Delta$Odds calculated from the logistic regression reveals that among galaxy \textit{clusters} (left panel of Figure \ref{fig:nuclboth}, i.e., A3558, A1736a, Coma, Virgo and Fornax) the rate at which galaxy nucleation varies with magnitude is similar within 5\%. This is similar to what was found in \citetalias{zanatta21}. However, for Local Volume samples the scenario is slightly different. The $\Delta$Odds calculated for C22 satellites of late-types is within less than 1\% of the average $\Delta$Odds calculated for galaxy clusters, but C22 satellites of early-types, on the other hand, present a $\Delta$Odds at least 10\% different to galaxy clusters but in agreement to what is observed for the H21 Local Volume sample. In fact, at magnitudes fainter than $M_I \sim -12$ mag, dwarf satellites of early-type galaxies at the Local Volume have a nucleation probability at least compatible but in average higher, within uncertainties, to that of galaxy clusters. \citet{hoyer21} have already identified a steeper slope in the nucleation fraction for the Local Volume relative to cluster environments using a logistic methodology\footnote{\citet{hoyer21} employed a Monte Carlo Markov Chain estimator to fit a \textit{logistic function}, with parameters $\kappa$ and $M_{50}$, for the slope and midpoint, respectively. This differs essentially from the methodology used in this work since we employ the logistic regression in the more straightforward GLM framework. Since these are two completely distinct approaches, their $\kappa$ and $M_{50}$ parameters are not immediately comparable to the $\Delta$Odds (or the $\beta_{2}$ slope) and $M_{I,f_{n,50}}$ presented in this work.}. In addition, the fact that we use in the present analysis two independent samples for the Local Volume further provides reliability to the difference in $\Delta$Odds observed this time (see Appendix \ref{appendixC}, where we show results for the nucleation fraction of the undivided C22 sample compared to that of the H21 sample. There are no significant discrepancies between these two samples for the entire magnitude range analysed). The increased nucleation for less luminous/massive dwarfs in the Local Volume in regards to galaxy clusters is complicated to be explained within the paradigm of the hierarchical growth of cosmological structures. It establishes that over time galaxy groups are incorporated into larger galaxy clusters \citep[][to name a few, and \citet{lisker18} for an example of the ongoing process of group accretion into the Virgo cluster]{white78, white91, kauffman93, navarro96}. Therefore, to explain that clusters such as Virgo present a lower nucleation fraction than Local Volume groups at the faint end, some process would be required to preferentially disrupt nucleated dwarfs in relation to their non-nucleated counterparts during the group accretion process. But this is in direct contradiction with the idea that nucleated objects are less likely to be disrupted by environmental effects in comparison to non-nucleated galaxies \citep[e.g.][]{leaman21}. Nevertheless, in all considered environments, from groups at the Local Volume to galaxy clusters, the modelled dependence of dwarf galaxy nucleation with luminosity is within a difference in $\Delta$Odds $\lesssim$ 10\%. Hence, we consider that the result observed in \citetalias{zanatta21}, i.e., that \textit{the rate at which galaxy nucleation vary with luminosity does not depend substantially on the environment.}, is consistent with the results observed in the current analysis.

\subsection{Nucleation Fraction dependence on environment}

As observed in Figure \ref{fig:nuclboth}, the higher estimated nucleation fraction observed among the environments studied in this work is that of the Coma and A3558 cluster, across the entire magnitude range analysed. On the other hand, the lowest nucleation fraction is found to be that estimated for dwarf satellites of Late-Type galaxies in the Local Volume C22 sample. In Figure \ref{fig:hostmass} we are able to more clearly verify the environment effect on galaxy nucleation: more massive environments present a higher probability of hosting nucleated galaxies, at fixed luminosity. This was previously observed in \citetalias{zanatta21} and in the present work we expand the analysis including the SSC clusters and the updated Local Volume and Fornax samples. The secondary dependence of galaxy nucleation on environment has already been observed in several works (\citealt{binggeli87, cote06, lisker07}; \citetalias{sanchez-janssen19b}; \citealt{hoyer21, carlsten22b, su22, fahrion22}). In an attempt to explain these results, \citetalias{sanchez-janssen19b} proposes a biased formation scenario for star clusters, similar to that proposed by \citet{peng08}, in which galaxies presently located in higher density environments experience an earlier onset of star formation sustaining higher star formation rates (SFR) and SFR surface densities. This would naturally lead to the formation of more massive star clusters as well as a larger mass fraction in star clusters in more dense environments, since early cluster formation is proportional to host galaxy mass with a near-universal efficiency \citep{kruijssen15}. 

\citet{leaman21} recently proposed a model for GC inspiral where the observed trends of the nucleation fraction can be reproduced by specific galaxy mass-size relations and the assumed GCMF. They argue that the increased nucleation fraction in more massive environments is possibly due to the disruption of most diffuse galaxies by the cluster potential. NSCs in diffuse galaxies are less likely to form, since their progenitor GCs are formed farther from the galaxy centre and not likely to be sufficiently massive to survive tidal disruption over the inspiral process. Compact dwarfs would naturally be more prone to form NSCs and be less susceptible to be disrupted by tidal effects in high density environments. \citet{marleau21} investigated a sample of 2210 ultra-diffuse galaxies (UDGs) in low-density environments in the nearby Universe ($z < 0.01$) and found no preferential mass-size relation for their sample in comparison to Virgo, Fornax and Coma dwarfs. \citet{su22} points that this evidentiates an important caveat of the scenario proposed by \citet{leaman21}: if diffuse galaxies are to be preferentially disrupted in more massive environments, then one would expect to see a higher fraction of such objects in low-density environments. It is possible, however, that deeper observations of low density environments will be able to reveal, in the future, a population of ultra-diffuse non-nucleated galaxies that is beyond the limiting magnitude of the currently available surveys. 

It is also well established in the literature that the fraction of nucleated galaxies tends to increase the closer objects are to the associated environment gravitational centre \citep{binggeli87,ferguson89,lisker07,ordenes-briceno18,hoyer21, carlsten22a, su22, fahrion22}. \citet{carlsten22a} argues as a possibility that the increased nucleation observed in galaxy clusters, when compared to Local Volume samples, can arise from the fact that observations of clusters are universally biased to central regions. This is the case for all clusters analysed in this work, with the exception of the Fornax sample which extends beyond the Fornax cluster virial radius \citep{su22}. \citet{hoyer21} analysed the matter in the Local Volume, as well as in the Virgo and Fornax clusters, with data from \citet{ferrarese20} and \citet{su21}, respectively, and did not find any significant difference in nucleation fraction due to galactocentric distance. Furthermore, \citet{su22} also investigated this issue for galaxies in the Fornax main cluster and the Fornax A group, finding a \textit{decrease} in nucleation with galactocentric distance in both environments. Hence, we consider the \textit{increase} in the nucleation fraction observed in more massive environments to be unlikely to arise from a galactocentric bias. But it is true that the addition of more radially complete datasets for large clusters, such as Virgo and Coma, would be an important inclusion to future studies of the environment influence on galaxy nucleation.    
Overall, it is clear from several observational results that nucleated dwarfs form a distinct and biased subpopulation. When compared to non-nucleated dwarfs, they present more circularised orbits \citep{ferguson89, lisker07, lisker09}, have higher axial ratios (\citealt{lisker07, eigenthaler18, venhola18}; \citetalias{sanchez-janssen19b}; \citealt{su22}), have more compact light profiles \citep{denbrok14}, host older stellar populations \citep{lisker08} and possibly feature higher GC mass fractions \citep{miller98, sanchez-janssen12} and higher GC abundances \citep{amorisco18}. Moreover, a better understanding of the biased growth of NSCs in denser environments can help constrain the star formation conditions in the early Universe. Regardless of the specific processes involved in NSC formation, they must be related to increased SFR which is significantly more common at earlier epochs \citep[and references therein]{madau14}. 

An interesting corollary of Fig. \ref{fig:hostmass}, is the possibility of using the nucleation fraction as an estimator for the halo mass of galaxy systems or clusters. The relation between GC and NSC growth previously stated would make it not too far-fetched to expect the NSC occupation fraction to be a viable estimator of environment halo mass, since it is well known that GC mass is related to the halo mass of their host galaxies \citep{spitler09}. For example, in this work we used for the A1736a cluster the halo mass estimate from \citet{lopes18} for the whole A1736 cluster, i.e., A1736a and A1736b, due to a lack of any estimate just for A1736a in the literature. Given the estimated $M_{I,f_{n,50}}$ of A1736a of $-15.01${\raisebox{0.5ex}{\small$_{2.30}^{1.06}$}} mag and the linear relation obtained from Fig. \ref{fig:hostmass}, one would expect a more accurate halo mass estimate for A1736a to be log$(M/M_{200}) = 12.54\pm9.76$. Despite the current uncertainties in this estimate, in the future, with nucleation data for additional environments, the relation between $M_{I,f_{n,50}}$ and environment halo mass might be an useful tool for extragalactic studies.  

\section{Summary and Conclusions}
\label{sec:conclusion}

Our main goal in this work is to investigate observational evidences of the environmental influence in the nucleation of galaxies. To this end, we use deep HST/ACS imaging in the F814W band to detect and analyse the nucleation fraction of quiescent dwarf galaxies in two galaxy clusters within the largest mass concentration in the nearby universe, the Shapley Supercluster. We then compare it with all other environments where nucleation data is available in the literature, enabling the analysis of the frequency of galaxy nucleation in host haloes from $5\times10^{14}$ in A3558 to $10^{12} \mathcal{M}_{\odot}$ at the Local Volume. This is motivated by the results of \citet{sanchez-janssen19b} and \citet{zanatta21} that showed that, for a fixed magnitude, galaxies are more likely to be nucleated in more massive environments. The Shapley clusters analysed in this work comprise the central and extremely dense Abell 3558 cluster as well as the northern and more modest, in terms of mass, Abell 1736a cluster. We find a total of 269 galaxies in the two clusters employing a dedicated algorithm combining \textsc{SourceExtractor} and \textsc{galfit} to detect galaxies between $-7.5 > M_I > 19.0$ mag and characterise their photometric and morphological properties. NSCs are identified by visual inspection and confirmed using 2D full image modelling. We then use Hierarchical Baysean Logistic Regression to model the dependence of nucleation fraction and galaxy absolute magnitude for the Shapley clusters and compare it to the same quantity estimated, using literature data, for quiescent dwarf galaxies in the Coma, Virgo and Fornax clusters as well as in groups and the field at the Local Volume. Our main conclusions are:

\begin{itemize}
    \item[1)] Abell 3558 has an estimated nucleation fraction, in the entire range of galaxy luminosity analysed, very similar to that of the Coma cluster. This is in agreement with these two clusters very similar estimated halo masses. A1736a, on the other hand, presents a nucleation fraction that is lower, within uncertainties, than any other galaxy clusters analysed and more compatible with Local Volume samples. 

    \item[2)] The rate at which galaxy nucleation fraction varies with magnitude is virtually universal, with brighter dwarfs being more likely to host NSCs. The mean difference in the rate of change with magnitude of the nucleation odds is 5\% among galaxy clusters and a maximum of 10\% in between galaxy clusters and Local Volume groups.  

    \item[3)] Including the data for Abell 3558 and Abell 1736a, as well as the most up-to-date literature data for nearby clusters and groups, we confirm the results of \citet{zanatta21} regarding the environment dependence on galaxy nucleation and its relation to the environment estimated halo mass. Galaxies in more massive environments have an increased likelihood to host NSCs. We find that the luminosity at which half of the dwarf galaxies contain an NSC is inversely proportional to the virial mass of the host halo, $L_{I,f_{n50}} \propto \mathcal{M}_{200}^{-0.2}$.  

\end{itemize}    

We have presented detections and photometry measurements for dwarf galaxies in two of the most extreme environments yet analysed in the context of galaxy nucleation. The comparison of their nucleation fraction with the one for galaxies in nearby clusters and groups evidentiates galaxy nucleation as a very complex process that involves both luminosity/mass and environment. Furthermore, our results present several important observational constraints to be taken into consideration in theories and models concerning the formation and growth of stellar nuclei.

We find it critical to raise the statistical significance of studies on the nucleation environmental dependence. First, by increasing the availability of nucleation data in other groups and clusters, in particular for massive clusters like Coma and A3558 and less massive haloes that the ones studied here. Secondly, by employing robust statistical methodologies avoiding outdated methodologies based on arbitrary binning that are easily biased by low number statistics or incomplete magnitude coverage. This is even more important now that recent and upcoming missions such as \textit{Euclid}, the \textit{Roman Space Telescope} or the The \textit{Chinese Space Station Telescope} allow us to investigate large NSC samples with deep imaging over large areas. These data will enable us to properly investigate also other important questions such as how galaxy morphology affects nucleation, in particular compactness and size, as predicted by models \citep{antonini15, mistani16, leaman21, fahrion22}. Finally, the full characterisation of the processes involved in NSC formation and growth cannot rely solely on occupation statistics. Studies on the stellar populations and kinematics of nuclei in different environments and for a wide range of galaxy masses are also invaluable for this investigation \citep{kacharov18, fahrion19, johnston20, carlsten22b}. 

\section*{Acknowledgements}
 
We thank the anonymous referee for the helpful insights and suggestions. This work is based on observations with the NASA/ESA \textit{Hubble Space Telescope}, obtained at the Space Telescope Science Institute, which is operated by AURA, Inc., under NASA contract NAS\,5-26555. These observations are associated with GO Programs \#10429. EZ acknowledges funding from the grant \#2022/00516-6, São Paulo Research Foundation (FAPESP). JPB was supported in part by the International Gemini Observatory, a program of NSF’s NOIRLab, managed by AURA, Inc., under a cooperative agreement with the National Science Foundation on behalf of the Gemini partnership.
 
\section*{Data availability}
 
The data underlying this article are available in the article and at the MAST HST archive (https://archive.stsci.edu/hst/) under the program ID 10429. Detection catalogues are available under request. 




\bibliographystyle{mnras}
\bibliography{ref.bib} 




\appendix
\section{References for the data used in this work not from the Shapley Supercluster}
\label{appendixA}

In this work we study the nucleation fraction in the A3558 and A1736a clusters and in other environments using literature data. In Table\,\ref{tab:appendixA} we list the other systems included in the analysis, namely the Coma, Virgo and Fornax clusters, as well as the Local volume ($D < 12$ Mpc)  the C22 and the H21 samples. In the table we indicate the number of quiescent satellites in each environment, as well as the literature sources for the photometry and the nucleation classification. 

Where applicable, we have converted the published magnitudes to the $I$-band. We adopt conversions based on the ones presented in \citet{blanton07}:

\begin{align}
    (B-i) \approx 1.10, \\
    (i - I) \approx 0.06.
\label{eq:conversions}
\end{align}

\begin{table*}
\caption{Source of the photometry and nucleation classification for the data used in this work. From left to right: First column in the Local Volume is to which sample galaxies associated with a given host, shown in the second column, are assigned to in this work, based on the sources of photometry and nucleation (fifth and sixth columns, respectively). The third and fourth columns in the Local Volume show the morphology of the central galaxy (ET for early-types and LT for late-types) and the number of objects part of the sample for each system. As there is no objective host classification in the H21 sample, we only show the number of objects and sources for photometry and nucleation. We refer the reader to \citet{hoyer21} and  \citet{hoyer23} for additional details. Finally, at the bottom we show information for the samples of nucleation in Galaxy clusters, from left to right: In the first column the galaxy cluster, in the second the number of objects in each sample and third and fourth columns the sources of photometry and nucleation, respectively. }
\label{tab:appendixA}

\resizebox{\textwidth}{!}{%
\begin{minipage}{\textwidth}
\centering
\begin{tabular}{ccccll}
\hline
\multicolumn{6}{c}{Local Volume}  \\                                              \\ 
\multicolumn{1}{c}{Sample}  & \multicolumn{1}{c}{Host}        & \multicolumn{1}{c}{Morph.}      & \multicolumn{1}{c}{N}   &  \multicolumn{1}{l}{Photometry Source}               &  \multicolumn{1}{l}{Nucleation  Source}                    \\ \hline
\rowcolor[HTML]{DDDDDD}
C22  &  NGC 1023       & Early-Type          & 15  & \citet{carlsten22b}              & \citet{carlsten22a}                    \\
C22  &  M104           & Early-Type          & 23  & \citet{carlsten20b}              & \citet{carlsten20a}                    \\
\rowcolor[HTML]{DDDDDD}
C22  &  NGC 5128 & Early-Type & 31 & \begin{tabular}[c]{@{}l@{}}\citet{carlsten20b};\\ \citet{carlsten22b};\\ \citet{muller19}\footnote{\citet{muller19} is the source photometry for the galaxies KK54 and KK58, and source nucleation for KK54.}\end{tabular} &
  \begin{tabular}[c]{@{}l@{}}\citet{carlsten20a};\\ \citet{carlsten22a};\\ \citet{muller19}; \\ \citet{fahrion20}\footnote{\citet{fahrion20} is the source nucleation for the galaxy KK58.}\end{tabular} \\
C22  &  NGC 3115 & Early-Type & 12 & \citet{carlsten22b}              & \citet{carlsten22a}                    \\
C22  &  NGC 3379 & Early-Type & 31 & \citet{carlsten22b}              & \citet{carlsten22a}                    \\
C22  &  NGC 4631 & Late-Type  & 7  & \citet{carlsten22b}              & \citet{carlsten22a}                    \\
\rowcolor[HTML]{DDDDDD}
C22  &  NGC 4565 & Late-Type & 16  & \citet{carlsten22b}              & \citet{carlsten22a}                    \\
C22  &  NGC 4258 & Late-Type & 9   & \citet{carlsten22b}              & \citet{carlsten22a}                    \\
\rowcolor[HTML]{DDDDDD}
C22  &  M51      & Late-Type & 4   & \citet{carlsten22b}              & \citet{carlsten22a}                    \\
C22  &  NGC 891  & Late-Type & 2   & \citet{carlsten22b}              & \citet{carlsten22a}                    \\
\rowcolor[HTML]{DDDDDD}
C22  &  NGC 6744 & Late-Type & 3   & \citet{carlsten22b}              & \citet{carlsten22a}                    \\
C22  &  NGC 628  & Late-Type & 6   & \citet{carlsten22b}              & \citet{carlsten22a}                    \\
\rowcolor[HTML]{DDDDDD}
C22  &  NGC 5457 & Late-Type & 2   & \citet{carlsten22b}              & \citet{carlsten22a}                    \\
C22  &  NGC 5236 & Late-Type & 4   & \citet{carlsten22b}              & \citet{carlsten22a}                    \\
\rowcolor[HTML]{DDDDDD}
C22  &  NGC 5055 & Late-Type & 6   & \citet{carlsten22b}              & \citet{carlsten22a}                    \\
C22  &  NGC 4826 & Late-Type & 3   & \citet{carlsten22b}              & \citet{carlsten22a}                    \\
\rowcolor[HTML]{DDDDDD}
C22  &  NGC 4631 & Late-Type & 1   & \citet{carlsten22b}              & \citet{carlsten22a}                    \\
C22  &  NGC 4736 & Late-Type & 1   & \citet{carlsten22b}              & \citet{carlsten22a}                    \\
\rowcolor[HTML]{DDDDDD}
C22  &  NGC 4517 & Late-Type & 4   & \citet{carlsten22b}              & \citet{carlsten22a}                    \\
C22  &  NGC 3627 & Late-Type & 7   & \citet{carlsten22b}              & \citet{carlsten22a}                    \\
\rowcolor[HTML]{DDDDDD}
C22  &  NGC 3521 & Late-Type & 2   & \citet{carlsten22b}              & \citet{carlsten22a}                    \\
\rowcolor[HTML]{DDDDDD}
C22  &  NGC 1291 & Late-Type & 4   & \citet{carlsten22b}              & \citet{carlsten22a}                    \\
\rowcolor[HTML]{DDDDDD}
C22  &  NGC 1808 & Late-Type & 3   & \citet{carlsten22b}              & \citet{carlsten22a}                    \\
C22  &  NGC 2903 & Late-Type & 3   & \citet{carlsten22b}              & \citet{carlsten22a}                    \\
\rowcolor[HTML]{DDDDDD}
C22  &  MW       & Late-Type & 22  & \begin{tabular}[c]{@{}l@{}}\citet{carlsten22b};\\ \citet{karachentsev13} \end{tabular}     & \begin{tabular}[c]{@{}l@{}}\citet{carlsten22a};\\ \citetalias{sanchez-janssen19b}\end{tabular}                  \\
C22  &  M81      & Late-Type & 40  & \begin{tabular}[c]{@{}l@{}}\citet{carlsten22b};\\ \citet{karachentsev13} \end{tabular}     & \begin{tabular}[c]{@{}l@{}}\citet{carlsten22a};\\ \citetalias{sanchez-janssen19b}\end{tabular}                 \\
\rowcolor[HTML]{DDDDDD}
C22  &  M31      & Late-Type & 60  & \begin{tabular}[c]{@{}l@{}}\citet{carlsten22b};\\ \citet{karachentsev13} \end{tabular}     & \begin{tabular}[c]{@{}l@{}}\citet{carlsten22a};\\ \citetalias{sanchez-janssen19b}\end{tabular}                 \\
H21  &  ---\footnote{No Host galaxy information is directly available in the H21 sample source catalogues, only a "main disturber" galaxy \citep{karachentsev13}. See text for details.}       & --- & 108 & \begin{tabular}[c]{@{}l@{}}\citet{karachentsev13};\\ \citet{hoyer21} \end{tabular}             & \citet{hoyer21}                    \\ \hline
\multicolumn{6}{c}{Galaxy Clusters}       \\                                       \\ 
\multicolumn{3}{l}{ID}            & \multicolumn{1}{l}{N}   &  \multicolumn{1}{l}{Photometry Source}                                                                         & \multicolumn{1}{l}{Nucleation Source} \\ \hline
\rowcolor[HTML]{DDDDDD}
\multicolumn{3}{l}{Abell 3559}    & 211 & This work                                                                                  & This work          \\
\multicolumn{3}{l}{Abell 1736a}   & 58  & This work                                                                                  & This work          \\
\rowcolor[HTML]{DDDDDD}
\multicolumn{3}{l}{Coma}          & 255 & \begin{tabular}[c]{@{}l@{}}\citetalias{zanatta21}; \citetalias{denbrok14} \end{tabular}    & \begin{tabular}[c]{@{}l@{}}\citetalias{zanatta21}; \citetalias{denbrok14} \end{tabular}           \\
\multicolumn{3}{l}{Virgo}         & 382 & \citetalias{sanchez-janssen19b}                                                            & \citetalias{sanchez-janssen19b}               \\
\rowcolor[HTML]{DDDDDD}
\multicolumn{3}{l}{Fornax}        & 263 & \citet{su22}                                                                               & \citet{su22}        
\end{tabular}%
\end{minipage}
}
\end{table*}

\section{Additional details from the hierarchical Bayesian logistic analysis}
\label{appendixB}

\begin{figure*}
    \centering
    \includegraphics[width=\linewidth]{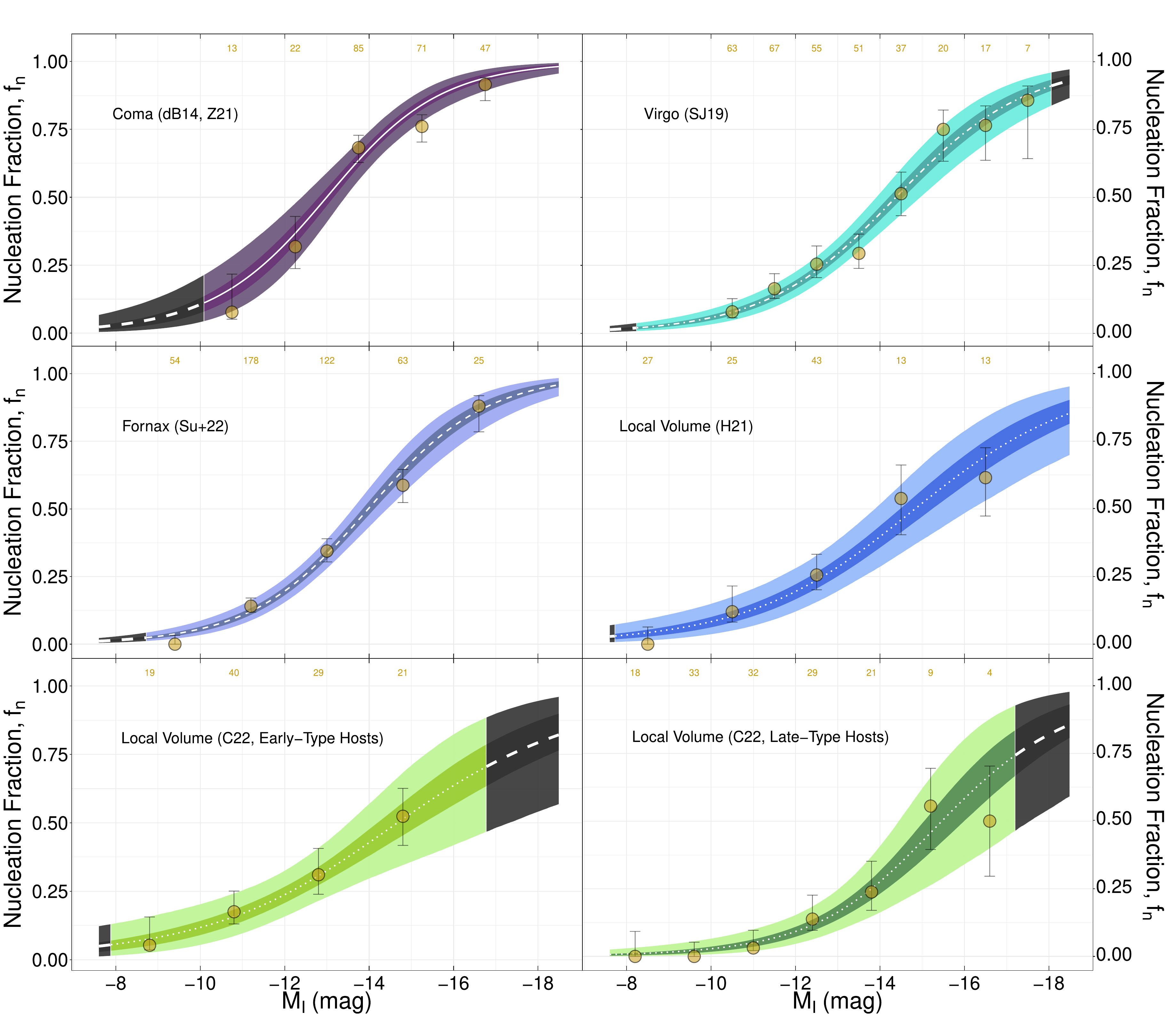}
    \caption{Same as Figure \ref{fig:nucl}, but for every other environment studied in this work, gathered from the literature references described in section \ref{sec:other} and Table \ref{tab:appendixA}. The resulting parameters from the logistic regression of these samples are shown in Table \ref{tab:model}.}
    \label{fig:individual}
\end{figure*}

In Figure \ref{fig:individual} we show the individual estimated nucleation fraction based on our methodology for every environment considered in Figure \ref{fig:nuclboth} other than the SSC. Similarly to what has been shown regarding A3558 and A1736a in Figure \ref{fig:nucl} and in \citetalias{zanatta21}, it is clear that the logistic regression follows closely the observed data while at the same time providing a smooth solution not dependent on arbitrary binning. This is specially important to properly compare datasets from different sources in a homogeneous way while still estimating uncertainties due to small sample sizes. 

We stress also the choice of magnitude range adopted in this work and the limitations of our methodology. It is well known that the full panorama of galaxy nucleation fraction deviates from the sigmoid curve of the standard logistic relation. Nucleation appears to universally peak around $\mathcal{M/M\odot} \sim 10^9$, declining for both higher and lower masses \citepalias{sanchez-janssen19b}. Therefore, to properly apply the logistic fit used in this work we are required to limit our analysis to the magnitude range where the nucleation fraction is monotonically crescent. A possibility to overcome this limitation, out of the scope of this work, is the employment of composite functions such as a logistic regression with splines or explore other functional forms within the GLM framework, such as Poisson regression. This is to be explored in future works if methods to detect precisely NSCs in large and bright galaxies are developed and catalogues in an extended magnitude range with nucleation information become more widely available.

\section{Additional remarks about the Local Volume samples}
\label{appendixC}

In this section we give further details on the process to collect the literature data used for Local Volume samples in this work, and then discuss additional remarks about the results of the nucleation fraction via logistic regression presented in Section \ref{sec:results}. 

The C22 sample is comprised of all galaxies used in \citetalias{zanatta21} to model the nucleation at the Local Volume with the addition of objects from the latest catalogue of the Exploration of Local VolumE Satellites (ELVES) Survey \citep{carlsten22b}. In such work, the final quiescent dwarf galaxy Local Volume sample of ELVES is presented, in continuation to the data presented in \citet{carlsten20b}. This sample is $\sim$ 50\% complete at $M_V < -9$ mag and $\mu_{0,V} < 26.5$ mag arcsec$^{-2}$ \citep{greene23}. In \citet{carlsten22a} the nucleation classification for galaxies in the ELVES sample was presented, but in such work objects are only associated with stellar masses, not photometric magnitudes. In order to solve this issue, we matched objects found in Tables 4 and 5 of \citet{carlsten22b} with the ELVES catalogue from \citet{carlsten22a}. A few objects present in \citet{carlsten22b} are not in the catalogue of \citet{carlsten22a}. These objects were not included in the ELVES catalogue due to distances to hosts being greater than the radial completeness limit of the survey (a projected radius of 200 kpc). However, for the statistical analysis of galaxy nucleation in \citet{carlsten22a} included additional objects given that a confirmed distance measurement was available (S. Carlsten, private communication). In order to have both photometric and nucleation information, in this work the C22 sample includes only objects that are both in \citet{carlsten22a} and \citet{carlsten22b}. Similarly to what was done in \citetalias{zanatta21}, we include in the C22 sample additional data from outside the ELVES catalogue, for further coverage of important hosts such as the MW, M81, M31 and NGC 5128. This is detailed in Table \ref{tab:appendixA}. 

The H21 Local Volume sample used in this work is based on \citet{hoyer21}, which is drawn from the Updated Nearby Galaxy Catalogue (UNGC) \citep{karachentsev13}. Despite not being as deep as the ELVES catalogue, the UNGC is by far the best current reference for Local Volume dwarfs, being constantly updated, with the version used in \citet{hoyer21} containing 1246 objects in late February 2021. Complementary nucleation classification of UNGC objects was presented in \citet{hoyer21} and \citet{hoyer23}, based on archival HST data. We selected only early-type dwarfs from \citet{hoyer21} and with available nucleation classification. A great portion of objects in such catalogue had missing nucleation information due to data for such objects not being available in archival HST data. This resulted in a final H21 sample of 170 galaxies within the magnitude limits of our analysis. One important caveat from the H21 sample is that it does not presents a reliable way to separate objects into dwarfs around early and late-type centrals, as it is done for C22. Following the analysis of \citet{karachentsev13} for the UNGC, for each object, instead of directly assigning central hosts, the \citet{hoyer21} catalogue indicates a list of closest interacting objects ("main disturbers", MD), which in several cases includes no central galaxies or multiples thereof. This is to be expected due to the claimed high fraction of isolated galaxies in the sample, but we also find that there are a number of objects in the \citet{hoyer21} catalogue which MDs are discrepant to the ones available for the same objects in the UNGC. Therefore, the analysis of the nucleation in groups regarding the morphology of central galaxies is performed only with the C22 sample.  

\begin{figure}
    \centering
    \includegraphics[width=\linewidth]{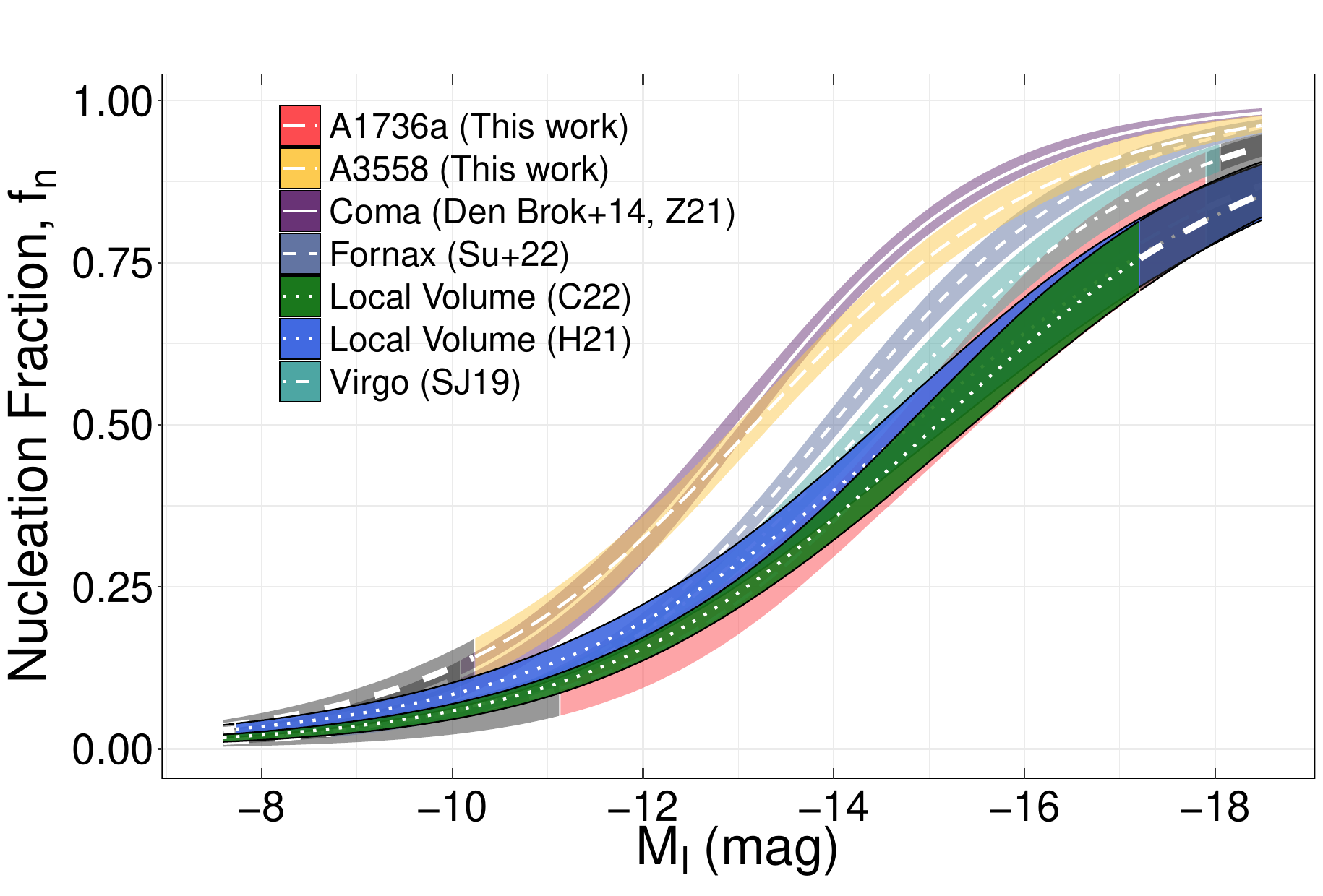}
    \caption{Same as Figure \ref{fig:nuclboth}, but for the logistic modelling run with the undivided C22 sample, shown as the dark green curve. Colours are the same as in Figure \ref{fig:nuclboth}. With exception of the H21 sample, all other environments are shown in the background for reference. The parameters resulting from the Baysean Logistic Regression run with these samples are shown in Table \ref{tab:model_divC22}.}
    \label{fig:divC22}
\end{figure}

Nevertheless, the reasoning behind including two independent Local Volume samples in this work is twofold: first, there is little overlap between the two samples, with as few as 54 objects in common between C22 and H21. In addition, based on the local environment density analysis performed in \citet{hoyer21}, around 35\% of objects in the H21 sample can be considered as "isolated"\footnote{The reader is suggested to read Section 3.4 of \citet{hoyer21} for additional details, but basically in this work we consider as likely isolated galaxies with a negative tidal index for the tenth most influential neighbouring galaxy, $\Theta_{10}$, as present in the \citet{hoyer21} catalogue.}. The C22 sample, in the other hand, is specifically focused on dwarf satellites of Local Volume galaxies. Therefore, the H21 sample comprises a large portion of isolated galaxies in the Local Volume not included in the C22 sample. Secondly, the observed increased nucleation fraction, relative to galaxy clusters, for objects in the C22 sample around Early-Type galaxies fainter than $M_I \sim -12$ mag is intriguing, as discussed in Section \ref{sec:discussion-lumi}. In order to mitigate the possibilities of a selection bias being the cause of this result, in Figure \ref{fig:divC22} and Table \ref{tab:model_divC22} we show the same logistic modelling described in Section \ref{sec:modelling} but considering an undivided (on the morphology of the central group galaxy) C22 sample. One can see the remarkable similarity between the C22 and H21 samples, despite each being based on two independent studies. It is however, still possible that some kind of bias is present in the C22 sample and affects preferentially the sub-sample of dwarf satellites of Early-Type galaxies, since the C22 sample of dwarfs around Late-Type galaxies shows compatible estimated values of $M_{I,f_{n,50}}$ and $\Delta$Odds with what is expected based on the analysis of the other environments in this work and \citetalias{zanatta21}. Nevertheless, we expect that missions such as \textit{Euclid} will provide wide-field photometry of Local Volume galaxies with unprecedented spatial coverage, mitigating any observational or selection bias that may be present in the current available data. On the other hand, if this result is unbiased, then it could indicate interesting aspects of the evolution of dwarf galaxies in the transitional group-to-cluster environment.   

\begin{table*}
\centering
\caption{Same as Table \ref{tab:model}, but for the logistic modelling presented in Figure \ref{fig:divC22}, where we include an undivided C22 sample.}

\label{tab:model_divC22}
\begin{tabular*}{\linewidth}{@{\extracolsep{\fill}}rrcrr}
\hline
 & $M_{I,f_{n50}}$ & $\Delta$Odds (\%) & $\beta_{1}$ (intercept) & $\beta_{2}$ (slope)\\\\
 \hline
\multicolumn{1}{l}{A3558}       & $-13.16${\raisebox{0.5ex}{\small$_{0.52}^{0.61}$}} & -46.95  & $ -8.35 \pm 1.56$ & $-0.63 \pm 0.115$ \\
\multicolumn{1}{l}{A1736a}      & $-14.99${\raisebox{0.5ex}{\small$_{2.22}^{1.04}$}} & -47.12  & $ -9.55 \pm 3.02$ & $-0.64 \pm 0.214$ \\
\multicolumn{1}{l}{Coma} & $-12.99${\raisebox{0.5ex}{\small$_{0.43}^{0.61}$}} & -52.27  & $ -9.61 \pm 1.54$ & $-0.74 \pm 0.110$ \\
\multicolumn{1}{l}{Virgo}            & $-14.37${\raisebox{0.5ex}{\small$_{0.58}^{0.45}$}} & -47.41  & $ -9.23 \pm 0.98$ & $-0.64 \pm 0.074$ \\
\multicolumn{1}{l}{Fornax}          & $-13.99${\raisebox{0.5ex}{\small$_{0.45}^{0.37}$}} & -51.13  & $-10.01 \pm 0.98$ & $-0.72 \pm 0.075$ \\
\multicolumn{1}{l}{Local Volume (H21)}      & $-14.83${\raisebox{0.5ex}{\small$_{1.38}^{0.95}$}} & -39.83  & $ -7.53 \pm 1.28$ & $-0.51 \pm 0.095$ \\
\multicolumn{1}{l}{Local Volume (C22)}      & $-15.08${\raisebox{0.5ex}{\small$_{1.32}^{0.81}$}} & -42.58  & $ -8.37 \pm 1.23$ & $-0.55 \pm 0.095$ \\
\hline
\end{tabular*}
\end{table*}


\bsp	
\label{lastpage}
\end{document}